\documentclass[a4paper,11pt]{article}

\usepackage{jcappub}
\usepackage{bm}
\usepackage{cleveref}
\usepackage{amsmath}
\usepackage{amssymb}
\usepackage{booktabs}    
\usepackage[table,xcdraw,dvipsnames]{xcolor}
\usepackage{collcell}
\usepackage{pgf}
\usepackage{aas_macros}
\usepackage{multirow}
\usepackage[most]{tcolorbox}

\newcommand{\gau}{^{(G)}}
\newcommand{\toph}{^{(t)}}
\subheader{IFT-UAM/CSIC-26-100}

\title{Weighted Webs: Morphology-Informed Marked Fields}

\author[a,b]{Mikel Martin Barandiaran,}
\author[c,d,e]{Jessica A. Cowell,}
\author[c]{David Alonso,}
\author[a]{Javier Carrón Duque}
\author[a,b]{and Juan García-Bellido}
\affiliation[a]{Instituto de Física Teórica UAM-CSIC,\\Campus de Cantoblanco, 28049 Madrid, Spain}
\affiliation[b]{Departamento de Física Teórica, Universidad Autónoma de Madrid,\\28049 Madrid, Spain}
\affiliation[c]{Department of Physics, University of Oxford,\\ Denys Wilkinson Building, Keble Road, Oxford OX1 3RH, United Kingdom}
\affiliation[d]{Kavli IPMU (WPI), UTIAS, The University of Tokyo,\\5-1-5 Kashiwanoha, Kashiwa, Chiba 277-8583, Japan}
\affiliation[e]{Center for Data-Driven Discovery, Kavli IPMU (WPI), UTIAS, The University of Tokyo,\\Kashiwa, Chiba 277-8583, Japan}

\emailAdd{mikel.martin@uam.es}

\abstract{The morphology of the cosmic web formed by the late-time matter distribution encodes cosmological information beyond that contained in standard two-point statistics. Marked power spectra provide a computationally efficient framework to access this higher-order information, by studying the two-point statistics of the density field ``marked'' (i.e. weighted) by a function of its local environmental density. In this work we explore the potential of marks that are sensitive to the morphology of this local environment, rather than simply its density.  We study a broad range of such marks, considering mark functions based on the smoothed density, tidal shear amplitude, local degree of isotropy and filamentarity, Gaussian transformations of these quantities and a local fractal dimension estimator. We quantify the merit of different marks in terms of their constraints on key cosmological parameters, including the matter abundance $\Omega_m$, the amplitude of fluctuations $\sigma_8$, and the mass of neutrinos $M_\nu$.
We have found that density-dependent marks continue to provide the largest improvements over the standard power spectrum, while morphology-based marks yield more modest improvements on their own. Nevertheless, combining density- and morphology-based marks consistently enhances cosmological constraints beyond what either class achieves separately, demonstrating that they probe complementary aspects of the underlying matter distribution. Within the family of shear-based morphology marks we find that the detailed functional form of the mark has only a minor impact on the resulting constraints, suggesting that the cosmological information is primarily encoded in the tidal shear field itself rather than on the particular non-linear transformation used to construct the mark. These results provide a systematic assessment of morphology-based marked statistics and explore which geometric properties of the cosmic web contribute most effectively to cosmological parameter inference. They also establish a physically motivated framework for future investigations of optimal marks and their perturbative connection to higher-order correlation functions.}

\begin{document}

\maketitle
\flushbottom

\section{Introduction}\label{sec:intro}
The large-scale structure (LSS) of the Universe is one of the richest cosmological probes. The distribution of matter overdensities on different scales as a function of time is sensitive to both the background expansion of the Universe and to the nature of the gravitational forces that drive their growth. At late times, the matter density field exhibits markedly non-Gaussian features, caused by the cumulative impact of non-linear gravitational collapse \cite{Zeldovich:1969sb}. For instance, the distribution of overdensities $\delta$ displays a heavy tail $\delta\gtrsim100$ dominated by dark matter haloes, virialised structures that dominate the total matter content \cite{Gunn:1972sv}. However, the space between haloes is filled by a rich hierarchy of structures, such as filaments, sheets, and voids, which form the so-called ``cosmic web'' \cite{Bond:1995yt,Springel:2006vs,0708.1441,Forero-Romero:2008wsm,Cautun:2014fwa,Libeskind:2017tun}. 
  
Recovering the primordial cosmological information that is now encoded in this non-Gaussian density field at late times is one of the most pressing problems for modern cosmology. This is a complex problem, since no single summary statistic can be shown to compress all information available in general non-Gaussian fields, as opposed to Gaussian data, where the correlation function or power spectrum are maximally informative observables. This has motivated the exploration of a very large variety of higher-order statistics (HOSs), aimed at extracting this non-Gaussian information. Examples include higher-order $N$-point correlation functions and polyspectra \citep{Scoccimarro:1997st,Bernardeau:2001qr}, Minkowski Functionals \citep{1994Mecke,1997Schmalzing}, persistent homology and Betti numbers \citep{Wilding:2020oza, Pranav:2016gwr}, wavelet scattering coefficients \cite{Cheng2020ANA, Valogiannis_2022}, void statistics \citep{pisani2019cosmic,Davies2020ConstrainingCW}, peak counts \citep{Fluri:2018fpg} and many others. Different HOSs are typically designed to balance and optimise three different aspects of cosmological inference: pure information content, the ability to produce accurate theoretical predictions for the cosmological models of interest, and the practicality of deploying these observables on realistic data, where the impact of large numbers of systematics must be characterised and mitigated.

In this context, marked power spectra are an attractive non-Gaussian observable for cosmology. In this approach, the density field is first ``marked'', applying spatially-varying weights to it defined in terms of properties of the density field itself. The two-point correlator of this marked field, as well as its cross-correlation with the original field, thus contain information encoded in higher-order correlators. Marked correlation functions were first formulated in \cite{1984Stoyan} in terms of marked point process, and early work employed them in the context of astrophysics, e.g. to study the luminosity dependence of galaxy clustering \citep{Beisbart_2000, Skibba_2006}, or the impact of environment in galaxy formation \cite{2009MNRAS.399..966S,2025arXiv250510429J,2021A&A...653A..35S}. In the context of cosmology, marked power spectra have been shown to be powerful observables, able to extract non-Gaussian information efficiently by e.g. up-weighting under-dense regions, where the impact of modifications to gravity and Dark Energy may be more prominent \citep{White:2016yhs, Armijo_2018, kärcher2024optimal, Valogiannis_2018}, or where mode-coupling of small-scale modes due to non-linear growth is suppressed \cite{1705.05328}. Their ability to extract significant cosmological information has been demonstrated in multiple works \cite{Massara_2023, massara2024sc}. Marked power spectra are also attractive in the context of data analysis due to their simplicity, relying on simple local weighting and existing power spectrum estimation methods, which makes them straightforward to apply to real data, both in the context of galaxy clustering \citep{Satpathy_2019,Armijo_2018,armijo2023new,massara2024sc} and, more recently, weak gravitational lensing \cite{Cowell:2025mov}. Finally, while the specific form of the mark function can be optimised \cite{Cowell:2024wyl}, choosing relatively simple marks (e.g. polynomial functions of the overdensity) makes it possible to develop accurate and reliable theoretical predictions from fundamental principles \cite{Philcox:2020fqx,2021Philcox,Marinucci_2024,Ebina:2024zkv,Ebina:2026qzf, 2026Marinucci}.

Most applications of marked power spectra have relied on mark functions that depend directly on the local value of the overdensity smoothed over a given scale, $\delta_R$. A priori this is a reasonable choice, as the local density has been shown to be the strongest indicator for halo abundances, clustering, and their scalar properties \cite{1406.4159,1801.04878}. However, the gravitational evolution of the cosmic web is in general governed by the second derivatives of the gravitational potential $T_{ij}\equiv\partial_i\partial_j\Phi$, and not just its Laplacian $\nabla^2\Phi\propto\delta$. Thus, important additional information may be recovered by studying the full tidal tensor. Indeed, tidal forces play a key role in the formation of the cosmic web \citep{0809.4135} (e.g. in the so-called ``T-web'' picture \cite{2310.03548}), defining the principal directions of gravitational collapse. It is therefore interesting to explore whether additional information may be gained by exploiting the local morphology of the cosmic web (as defined by the tidal field or otherwise) in the design of mark functions. This idea has been previously exploited in the literature, typically by dividing the field into different environment classes (void, filaments, sheets, and clusters) as determined by the local tidal forces \citep{kärcher2024optimal,2601.05934}.

In this work, we explore the possibility of exploiting the local morphology of the cosmic web within the framework of marked power spectra to extract information beyond that accessible using standard density-dependent mark functions. In particular, we will study the possibility of using functions dependent on the local shear tensor with varying levels of complexity, as well as the other quantifiers of the local dimensionality of the cosmic web, and compare the results against the information recovered using standard density-dependent marks. The paper is structured as follows: Section \ref{sec:theory} introduces the theoretical background between different approaches to describe the cosmic web, as well as the marked power spectrum framework. Section \ref{sec:methodology} then describes the procedure used to quantify the information content of different cosmic web mark functions. The results of this analysis are presented and discussed in Section \ref{sec:results}. We then conclude in Section \ref{sec:conclusions}.
  
\section{Theory}\label{sec:theory}

In this Section we introduce the theoretical ingredients underlying our morphological marked fields. We begin by reviewing the tidal tensor description of the cosmic web and the associated scalar invariants that characterise anisotropic gravitational collapse. We then discuss the local fractal dimension as an alternative geometric characterisation of the cosmic web. Finally, we construct marked fields based on these quantities and motivate their use as probes of cosmological information beyond standard two-point statistics of the density field.

\subsection{The Cosmic Web}\label{ssec:cosmic_web}

Let $\delta(\bm{x})$ denote the matter overdensity and let $\delta_R(\bm{x})$ be
the same field smoothed on a comoving scale $R$ with a Gaussian kernel. This overdensity sources a local peculiar gravitational potential $\Phi(\bm x)$, which ultimately drives the anisotropic collapse of the cosmic web. The tidal tensor is the Hessian of this gravitational potential,
\begin{equation}\label{eqn:tidal_tensor_def}
    T_{ij}(\bm{x})=\partial_ i\partial_j\Phi(\bm{x})=\int\frac{d^3\bm{k}}{(2\pi)^3}\frac{k_i k_j}{k^2}\delta(\bm{k})\,e^{i\bm{kx}}\,,
\end{equation}
and its traceless part is commonly named the \textit{shear tensor}:
\begin{equation}\label{eqn:shear_tensor_def}
    s_{ij}(\bm{x})=T_{ij}(\bm{x})-\frac{T^k_k(\bm{x})}{3}\delta_{ij}=\int\frac{d^3\bm{k}}{(2\pi)^3}\underbrace{\left(\frac{k_i k_j}{k^2}-\frac{1}{3}\delta_{ij}\right)\delta(\bm{k})}_{s_{ij}(\bm{k})}\,e^{i\bm{kx}}\,,
\end{equation}
where $\delta_{ij}$ is the Kronecker delta. The above definitions set the Fourier convention that we adopt throughout this work.

According to the Equivalence Principle, a locally uniform gravitational field can always be ``gauged-away'' by transforming to a freely falling frame. Therefore, gravitational collapse cannot depend directly on the gravitational potential $\Phi$ or on its first derivatives, but only on tidal forces encoded in the second derivatives $\partial_i\partial_j\Phi$. The tidal tensor therefore provides a natural local descriptor of anisotropic structure formation via gravity, and of the morphology of the cosmic web \citep{1970Ap......6..320D,astro-ph/9507024,Forero-Romero:2008wsm,2310.03548}. 

The standard Poisson equation relates the comoving density field to the physical gravitational potential via $\nabla^2\Phi=4\pi G\bar\rho a^2\delta$. However, by absorbing these cosmological factors\footnote{We are already doing so implicitly in the definition given in \Cref{eqn:tidal_tensor_def}} into a rescaled potential $\Phi$, we can say that the density field is equal to the trace of the tidal tensor
\begin{equation}
    \delta(\bm{x})= \nabla^2\Phi(\bm{x}) =T_i^i(\bm{x})\,.
\end{equation}
As a $3\times 3$ tensor field, the tidal tensor defines not only one but three natural scalar invariants at each point in space. A simple representation of these invariants is comprised by the eigenvalues of $T_{ij}$. We denote these by $(\lambda_1,\lambda_2,\lambda_3)$, and assume, without loss of generality, the ordering $\lambda_1\geq\lambda_2\geq\lambda_3$. Another way of encoding the same information is through the coefficients $(I_1,I_2,I_3)$ of the characteristic polynomial of $T_{ij}$, related to the eigenvalues via
\begin{equation}\label{eqn:I_invariant_def}
        I_1=\lambda_1+\lambda_2+\lambda_3\,,\hspace{8mm}I_2 =\lambda_1\lambda_2+\lambda_1\lambda_3+\lambda_2\lambda_3\,,\hspace{8mm}I_3=\lambda_1\lambda_2\lambda_3\,,
\end{equation}
which possess the advantage of being numerically simpler and faster to evaluate. Finally, from a physical perspective the natural representation of the tidal invariants is via the density $\delta$ and the \textit{shear invariants} $(s^2,s^3)$, defined as
\begin{equation}\label{eqn:tidal_invariants_def}
\delta=T_i^i\,,\quad s^2=s_{ij}s^{ij}\,,\quad s^3=s_{ij}s_{jk}s_{ki}\,.    
\end{equation}
These quantities correspond respectively to first-, second-, and third-order invariants of the tidal eigenvalues, scaling schematically as $\delta=\mathcal O(\lambda)$, $s^2=\mathcal O(\lambda^2)$, and $s^3=\mathcal O(\lambda^3)$. Throughout this work we define the \emph{shear amplitude}
\begin{equation}
s\equiv\sqrt{s^2},
\end{equation}
which has the same physical dimensions as $\delta$ and the tidal eigenvalues. Likewise, $(s^3)^{1/3}$ also has the same dimensions, but represents a different quantity and should not be confused with $s$.\footnote{Since $s^2$ and $s^3$ contain different combinations of the tidal eigenvalues, one generally has $s\neq(s^3)^{1/3}$. Both are valid scalar descriptors of the tidal field, but encode different aspects of the local morphology.} The quantity $s^2$ is often simply called the shear invariant, while
$s^3$ receives names such as the \emph{cubic shear invariant} or, more commonly in the cosmology literature, \emph{prolateness parameter} \cite{Bardeen:1985tr,Desjacques:2007zg}
The shear is related to the tidal eigenvalues via
\begin{equation}\label{eqn:shear_eigval_sum}
    s^2\propto \sum_{ij}(\lambda_i-\lambda_j)^2\,,
\end{equation}
which highlights its geometric interpretation as an indicator of anisotropic stress. In fact, $s^2$ only vanishes for perfectly isotropic configurations, for which $\lambda_1=\lambda_2=\lambda_3$. The shear invariant also plays a central role in perturbative descriptions of large-scale structure, where it appears as one of the lowest-order non-local operators in Eulerian bias and effective field theory (EFT) expansions \cite{McDonald:2009dh,Desjacques:2016bnm}.

We emphasise that any function that can be built from any of the three triplet of scalars can trivially be regarded as a function of any other triplet: 
\begin{equation}\label{eqn:equivalent_scalars}
        (\lambda_1,\lambda_2,\lambda_3)\longleftrightarrow(I_1,I_2,I_3)\longleftrightarrow(\delta,s^2,s^3)\,.
\end{equation}
Using one or the other will be more convenient depending on the situation. It is worth noting, however, that the transformations relating each triplet with any other are non-linear.

\begin{figure}[htbp]
    \centering
    \includegraphics[width=\linewidth]{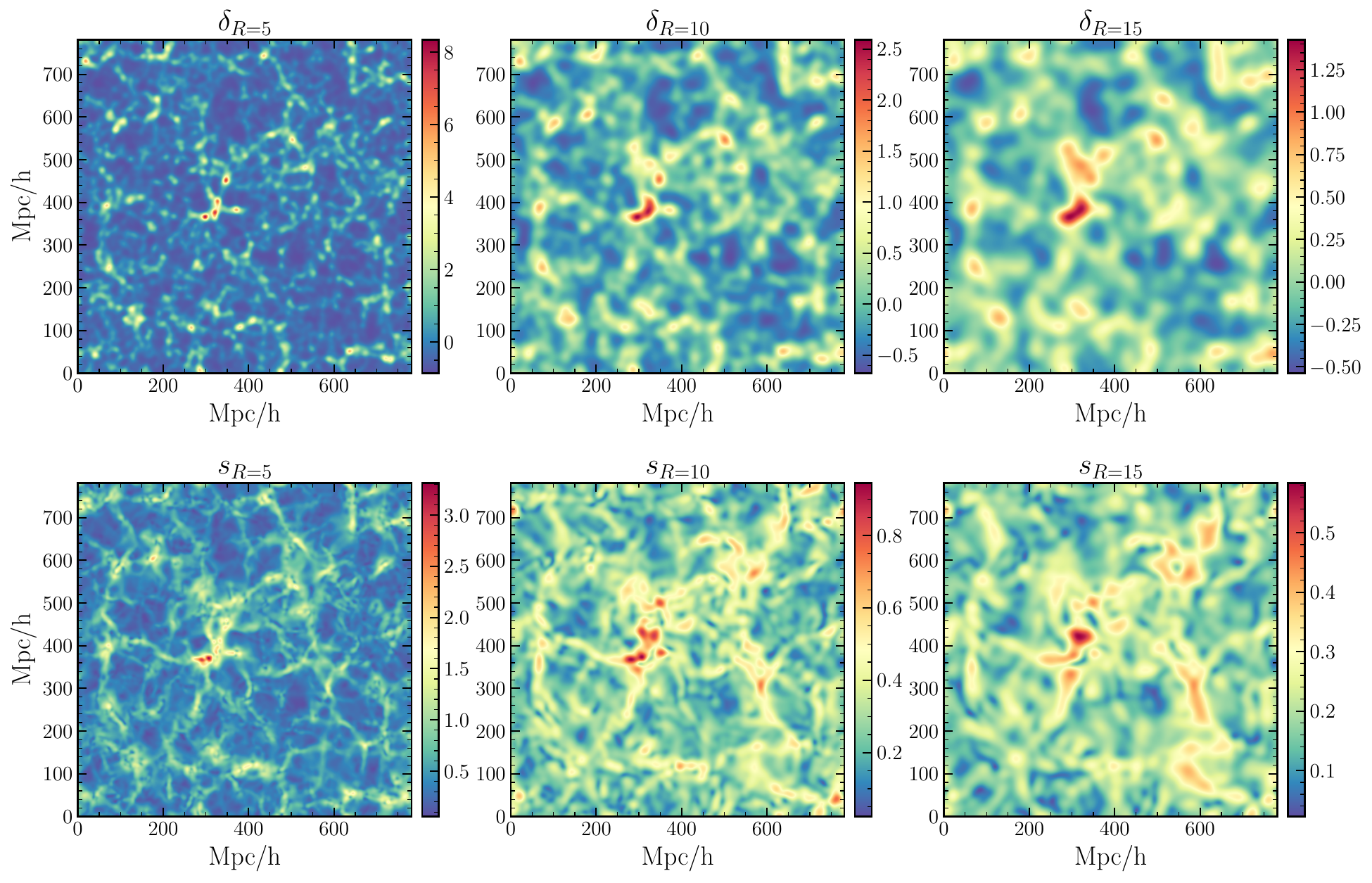}
    \caption{Two-dimensional slices of the smoothed density and shear fields, smoothed at comoving scales of $R=5,10,15$ $h^{-1}$Mpc.}
    \label{fig:delta_and_shear_slices}
\end{figure}

\Cref{fig:delta_and_shear_slices} shows two-dimensional slices of the smoothed density and shear-amplitude fields in one of the simulations used in \Cref{sec:methodology} for three representative smoothing scales. While the density field highlights the locations of overdense structures, the shear field emphasises regions where the gravitational collapse is strongly anisotropic (like filaments). Cluster interiors tend to exhibit relatively small shear owing to their nearly isotropic collapse, whereas filaments and walls stand out as elongated regions of large shear. Increasing the smoothing scale progressively suppresses small-scale features while preserving the large-scale morphology of the cosmic web.

Beyond providing a compact algebraic description of the local gravitational field, the tidal invariants carry direct information about the morphology of the surrounding structure. For this reason they are widely used to classify the cosmic web into distinct morphological environments, most notably through the tidal-tensor or \emph{T-web} classification \cite{Hahn:2006mk,0809.4135}. The most common approach is based on the eigenvalues of the tidal tensor, which quantify the strength of gravitational collapse or expansion along the three principal axes. By comparing each eigenvalue to a threshold $\lambda_{\rm th}$, one effectively counts the number of directions undergoing gravitational collapse. Although the Zeldovich approximation would naturally suggest $\lambda_{\rm th}=0$, numerical studies have shown that adopting a small positive threshold produces a classification that better matches the visually identifiable cosmic web \cite{0809.4135}. Physically, this reflects the fact that collapse along a given direction is not instantaneous once an eigenvalue becomes positive, but requires a finite amount of non-linear evolution. Typical values used in the literature lie in the range $\lambda_{\rm th}\simeq0.1$--$0.4$, depending on the smoothing scale. 

Regions collapsing along all three principal axes correspond to clusters, while collapse along fewer axes gives rise to progressively lower-dimensional structures such as filaments and walls. Conversely, regions with no collapsing directions are identified as voids. This leads to the following classification:
\begin{align*}
\text{Clusters:}   &\qquad \lambda_1,\lambda_2,\lambda_3 \ge \lambda_{\rm th},\\
\text{Filaments:}  &\qquad \lambda_1,\lambda_2 \ge \lambda_{\rm th}\;\text{ and }
                    \lambda_3 < \lambda_{\rm th},\\
\text{Walls:}      &\qquad \lambda_1 \ge \lambda_{\rm th}\;\text{ and }
                    \lambda_2,\lambda_3 < \lambda_{\rm th},\\
\text{Voids:}      &\qquad \lambda_1,\lambda_2,\lambda_3 < \lambda_{\rm th}.
\end{align*}
These environments are typically associated with progressively lower densities,
from clusters to voids, although the classification itself depends only on the
number of eigenvalues above the threshold $\lambda_{\rm th}$.

Rather than assigning each point to a single discrete environment through such threshold-based schemes, in this work we retain the full continuous information encoded in the tidal invariants and use it directly to construct the marked fields introduced in \Cref{ssec:marked_fields}. 

\subsection{The Local Fractal Dimension}\label{ssec:frac_dim}

An alternative characterisation of the cosmic web can be obtained from the local fractal dimension (LFD), which quantifies how the total mass enclosed around a point scales with radius. Intuitively, cluster-like regions behave as compact structures, meaning that the total mass does not significantly increase with radius, and therefore have $\rm{LFD}\sim 0$. Filaments resemble one-dimensional objects, so the mass increases roughly linearly with radius ($\rm{LFD}\sim 1$), and sheets are two-dimensional ($\rm{LFD}\sim 2$). Voids, with their sparsely distributed galaxies, have $\rm{LFD}\sim 3$. The local fractal dimension therefore provides a geometric probe of environmental morphology that is complementary to tidal classifications. We can thus use the LFD as another metric of the local morphology of the cosmic web, and test its ability to recover non-Gaussian information as a mark function.

\begin{figure}[htbp]
    \centering
    \includegraphics[width=\linewidth]{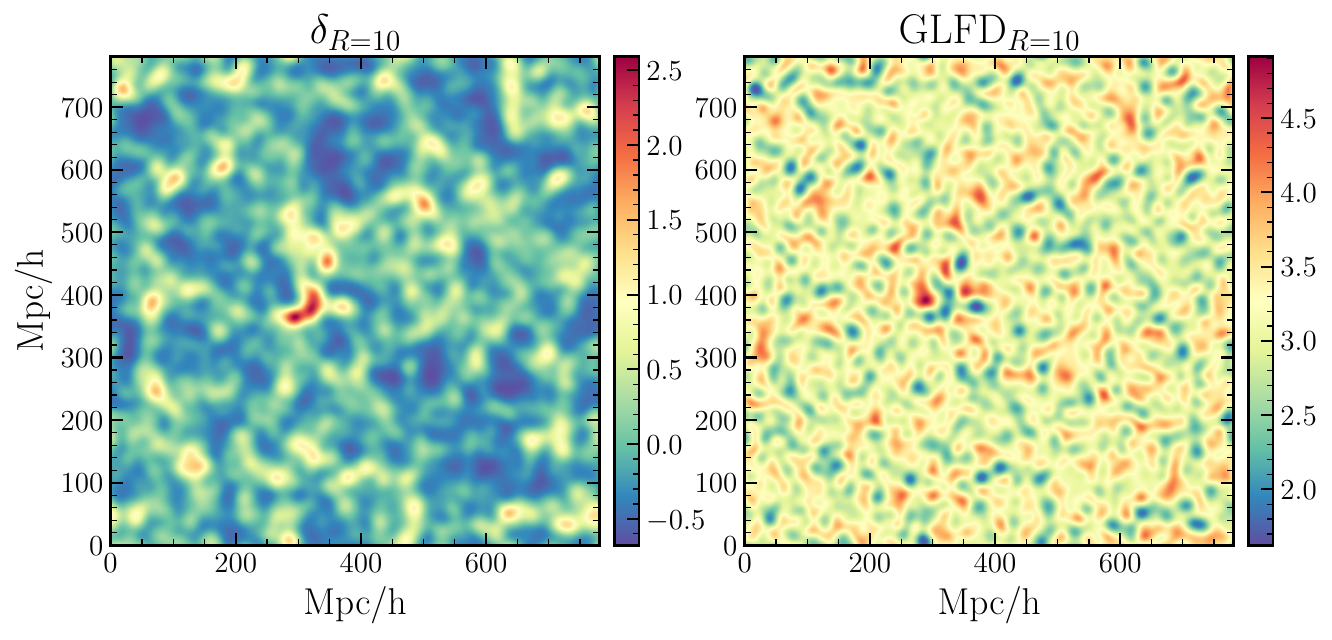}
    \caption{Two-dimensional slices of the smoothed density field (left) and the Gaussian Local Fractal Dimension (right), both computed with a Gaussian smoothing scale of $R=10\,h^{-1}\mathrm{Mpc}$. The realisation that produced this figure is the same as the one in \Cref{fig:delta_and_shear_slices}.}
    \label{fig:LFD_slice}
\end{figure}

In order to formally define the LFD, we start by computing  the total mass enclosed within a radius $R$ around a point $\bm{x}$ as:
\begin{equation}
    M_{<R}(\bm{x})=\int d^{3}\bm{y}[1+\delta(\bm{x}+\bm{y})]W_{R}(\bm{y}),
\end{equation} 
where $W_{R}(\bm{y})$ is a radially symmetric unnormalised window function, i.e., the volume integral of the window function is not 1, but some quantity $V(R)\propto R^{3}$. We will consider two types of window functions: the top-hat $(t)$, and the Gaussian $(G)$:
\begin{align}
    W_{R}\toph(\bm{x})=\Theta(x<R)&, \quad V\toph(R)=\frac{4\pi R^{3}}{3}    \label{eq:tophat_volume}, \\
    W_{R}\gau(\bm{x})=e^{-x^{2}/2}&, \quad V\gau(R)=(2\pi)^{3/2}R^{3},   \label{eq:gauss_volume}
\end{align}
where $\Theta(x<R)$ is $1$ for $x<R$ and $0$ otherwise. We will label the normalised window function $w_{R}(\bm{x})\equiv W_{R}(\bm{x})/V(R)$. It will also be useful to define the Fourier transform $\tilde{w}(\bm{k}R)$ of these functions. For the two types of windows above:
\begin{align}
    \tilde{w}\toph(\bm{k}R)&=\frac{3}{(kR)^{3}}(\sin kR - kR \cos kR)=3\,\frac{j_1(kR)}{kR}, \\
    \tilde{w}\gau(\bm{k}R)&=e^{-(kR)^{2}/2}.
\end{align}
Using these, the local enclosed mass can be written as:
\begin{equation}\label{eq:enclosed_mass}
    M_{<R}(\bm{x})\equiv V(R)[1+\delta_{R}(\bm{x})]=V(R)\left[1+\int\frac{d^{3}\bm{k}}{(2\pi)^{3}}e^{i\bm{k}\bm{x}}\delta_{\bm{k}}\tilde{w}(\bm{k}R)\right].
\end{equation} 
The LFD at a radius $R$, which we denote as $D(R)$, can be computed from $M_{<R}$ as:
\begin{equation}
    D_{R}(\bm{x})\equiv\frac{\mathrm{d} \log M_{<R}(\bm{x})}{\mathrm{d} \log R} .
\end{equation}
Using Eq. \ref{eq:enclosed_mass}, we can then write $D_{R}$ as:
\begin{equation}\label{eqn:LFD_def_as_ratio}
    D_{R}(\bm{x})=\frac{3+\delta_{\partial R}(\bm{x})}{1+\delta_{R}(\bm{x})},
\end{equation} 
where we have defined a modified smoothed field and its window function as:
\begin{align}
    \delta_{\partial R}(\bm{x})&\equiv\int\frac{d^{3}k}{(2\pi)^{3}}e^{ik\bm{x}}\delta_{k}\tilde{w}_{\partial}(kR), \\
    \tilde{w}_{\partial}(kR)&=3\tilde{w}(kR)+\frac{\mathrm{d}\tilde{w}(kR)}{\mathrm{d} \log(kR)}.
\end{align}
For the considered top-hat and Gaussian window functions, these take the following form:
\begin{align}
    \tilde{w}_{\partial}\toph(kR)=3\frac{\sin kR}{kR}, \hspace{12pt}
    \tilde{w}_{\partial}\gau(kR)=[3-(kR)^{2}]e^{-(kR)^{2}/2}.
\end{align} 
We note that, in both cases, for a realistic power spectrum, one can show that  $D_R(\bm{x})\rightarrow3\,$ as $R\rightarrow\infty$, which corresponds to the common intuition that the field is uniform at large enough scales. \Cref{fig:LFD_slice} shows a slice of a smoothed density field and the corresponding Gaussian LFD of the same smoothing scale. Upon visual inspection, it is clear to see that these fields are anticorrelated.

\begin{figure}[htbp]
    \centering
    \includegraphics[width=0.8\linewidth]{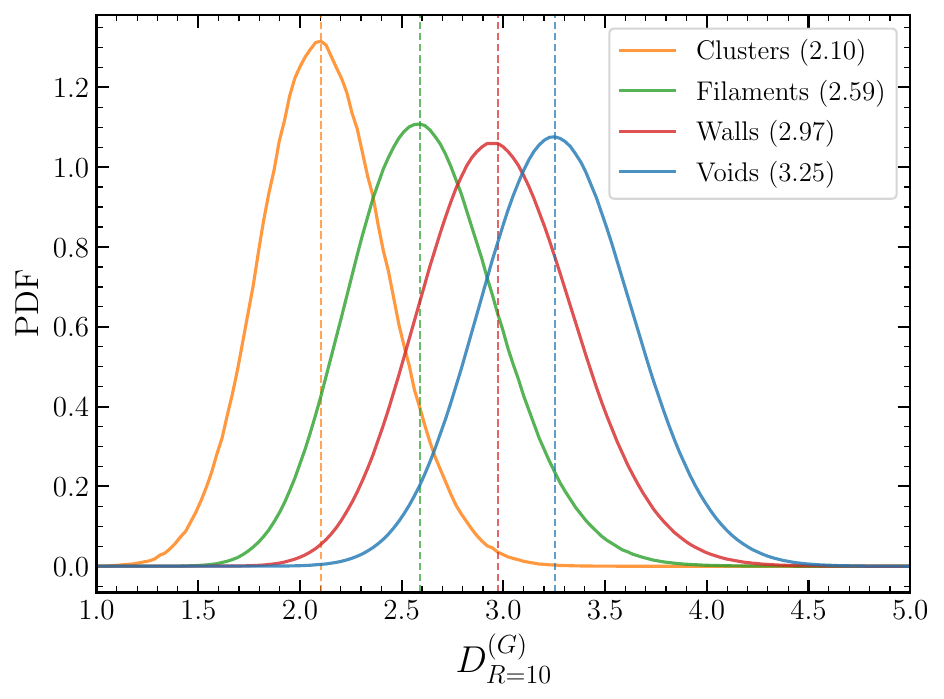}
    \caption{Empirical probability distribution function of the Gaussian local fractal dimension measured at $R=10$ $h^{-1}$Mpc, for the four different T-web environments. The environments were defined with the same smoothing scale and $\lambda_{\rm th}=0.2$. The vertical dashed lines indicate the corresponding mean values, which are also quoted in the legend.}
    \label{fig:LFD_histograms}
\end{figure}

To illustrate how the LFD relates to the traditional tidal classification, \Cref{fig:LFD_histograms} shows the probability distribution of the Gaussian LFD for voxels belonging to each T-web environment, where the latter is defined using the same smoothing scale $R=10\,h^{-1}\text{Mpc}$ and a threshold value of $\lambda_{\rm th}=0.2$.
Although the distributions overlap substantially, they exhibit a clear ordering: cluster environments preferentially populate the lowest values of $D_R$, followed by filaments, walls, and finally voids, whose distribution peaks closest to $D_R\simeq3$. This demonstrates that the LFD captures meaningful morphological information while remaining a continuous descriptor rather than a discrete environmental label.

The distributions are nevertheless shifted towards larger values than one might expect from the idealized picture in which clusters, filaments, walls and voids correspond exactly to dimensions $0$, $1$, $2$ and $3$, respectively. There are several reasons for this. First, the LFD measures the logarithmic growth of the enclosed mass around a point, and not the intrinsic dimensionality of an isolated structure, so nearby overdense regions can make the local mass-radius relation steeper and even produce values $D_R>3$. For instance, an underdense region sitting at a distance $R$ of a large overdensity can easily see its enclosed mass grow faster than $R^3$. Second, the Gaussian smoothing scale adopted throughout this work, $R=10\,h^{-1}{\rm Mpc}$, is comparable to or larger than the characteristic thickness of walls and to the transverse size of many filaments. As a result, the enclosed mass is substantially contaminated by neighbouring structures, causing filamentary and wall environments to appear more ``volume-filling'' than their idealized one- and two-dimensional counterparts. Consequently, the measured LFD should be interpreted as a continuous environmental measure rather than as a direct estimation of the intrinsic geometric dimensionality of the underlying cosmic-web structures.

\subsection{Marked Fields}\label{ssec:marked_fields}

Marked fields are constructed by assigning a spatially varying weight, or \emph{mark}, to every point in a field before computing its clustering statistics. Originally introduced in spatial statistics \cite{1984Stoyan} and later applied to galaxy clustering \cite{Skibba_2006}, the choice of mark is typically motivated by a physical property of interest. In cosmology, marks depending on the local density have received particular attention because they act as non-linear transformations of the density field, allowing two-point statistics of the marked field to recover part of the cosmological information encoded in higher-order correlations \cite{Philcox:2020fqx}. Marked statistics therefore provide a computationally efficient way of probing the non-Gaussian information generated by gravitational evolution while retaining the simplicity of power-spectrum estimators. The density dependent marked fields are commonly defined as 
\begin{equation}
    \Delta(\bm{x})=m(\delta_R({\bm{x}}))(1+\delta(\bm{x}))-\langle m(\delta_R({\bm{x}}))(1+\delta(\bm{x}))\rangle\,.
\end{equation}
Some authors additionally normalise the marked field by its mean value. However, such an affine transformation does not modify the Fisher information carried by the resulting observables \cite{Cowell:2024wyl}, so we adopt the simpler definition above throughout this work.

Existing marked statistics typically construct the mark as a function of the smoothed density alone. Here we instead promote the mark to depend on all the tidal invariants: 
\begin{equation}\label{eqn:tau_def}
    \tau(\bm{x})=m(\bm x)(1+\delta(\bm{x}))-\langle m(\bm x)(1+\delta(\bm{x}))\rangle\,,\quad\text{where }m(\bm x)=m(\delta_R(\bm x),s_R^2(\bm x),s_R^3(\bm x))\,.
\end{equation}
This generalization allows the mark to respond not only to overdensity but also to the anisotropy of the surrounding gravitational field, thereby incorporating explicit morphological information. It should be emphasised that this construction does not introduce new information beyond that already present in the density field: the tidal tensor is itself obtained from $\delta$ through \Cref{eqn:tidal_tensor_def}. Rather, the advantage of tidal marks lies in their ability to summarise information about the local morphology/anisotropy of the matter distribution, beyond its environmental density.

The complete workflow adopted in this paper is summarised schematically as 
\begin{tcolorbox}[
    enhanced,
    colback=blue!3,
    colframe=blue!50!black,
    boxrule=0.8pt,
    arc=2mm,
    halign=center,
    top=1mm,
    bottom=1mm,
]
$
\delta\rightarrow\delta_R\rightarrow T_{ij}\rightarrow(\delta_R,s_R^2,s_R^3)\rightarrow
m(\delta_R,s_R^2,s_R^3)\rightarrow\tau\rightarrow\{P_{\tau\tau},P_{\delta\tau},\ldots\}.
$
\end{tcolorbox}

Finding ``optimal'' mark functions has been a recurring goal in the marked-statistics literature \cite{White:2016yhs,Massara_2023,Beisbart_2000, Cowell:2024wyl}. One of the most studied ones is the White mark \cite{White:2016yhs}, which is defined as
\begin{equation}
    \mathrm{Wh}(\bm{x};R,p,\delta_s)=\left[1+\frac{\delta_R(\bm{x})}{1+\delta_s}\right]^{-p}\,.
\end{equation}
%
where the parameters $p$ and $\delta_s$ control the strength and characteristic density scale of the weighting, respectively. Positive values of $p$ upweight underdense regions with $\delta_R\rightarrow -1$, while it gives very little weight to regions with $\delta_R\gtrsim \delta_s$. For a smoothing scale of $R=10\,h^{-1}{\rm Mpc}$ the parameter values $(p,\delta_s)=(2,0.25)$ were found to be close to optimal in terms of constraining power \cite{White:2016yhs, Massara_2023}. Note however, that \citep{Cowell:2024wyl} showed large degeneracies in the information content when varying parameters of this mark. The top left panel of \Cref{fig:2D_hists} shows the functional form of such a mark.

Although traditionally the success of the White mark has been attributed to the fact that it upweights voids and underdense regions while zeroing out dense environments like clusters, Ref.~\cite{Cowell:2024wyl} showed that the cosmological information carried by marked statistics is invariant under arbitrary affine transformations of the mark function. Consequently, the absolute normalisation or even whether the mark is larger in voids than in clusters cannot by itself explain its performance, and a more profound study of how a certain mark reorganizes cosmological information is needed to pin-point the source of its ability to pull non-Gaussian information \cite{Philcox:2020fqx}. 

\subsubsection{Morphology-inspired marks}\label{sssec:morpho_marks}

The formalism introduced above allows the mark to depend on any scalar function of the local tidal tensor. Motivated by the different morphological environments of the cosmic web, we consider several families of marks that selectively enhance isotropic regions, filamentary structures, or a preferred range of shear values. All marks are constructed from fields smoothed on a scale $R$, which therefore acts as an additional hyperparameter controlling the environmental scale to which the mark is sensitive.\footnote{We further discuss the importance of the smoothing scale in \Cref{ssec:smoothing_scale}.}

In principle, one may construct a mark that depends on any combination of the tidal invariants $(\delta,s^2,s^3)$. In this work, however, we restrict ourselves to functions of the density $\delta$ and the quadratic shear invariant $s^2$. This choice is motivated both by their simple geometric interpretation and by their prominent role in perturbative descriptions of large-scale structure. We also explored the possibility of using as mark function the cubic invariant $s^3$, but found that it consistently yielded smaller error-improvement factors (see \Cref{sssec:eif}) than the $s^2$ counterparts. We therefore leave a more systematic exploration of cubic-invariant marks to future work.

The following list contains the family of marks that we have considered in this work. 

\begin{itemize}

\item \textbf{Analytically tractable marks.} In Ref.~\cite{Ebina:2024zkv} the smoothed density field $\delta_R(\bm x)$ was used as mark function. Although this choice is not the best performing mark in terms of constraining power, the associated marked power spectra may be described using perturbative analytical methods, which in turn allows for interpreting the possible origin of information gain and constraining power. In the same spirit, the first morphological mark we will study is the shear amplitude field $s_R(\bm x)=\sqrt{s_R^2(\bm x)}$. We choose $s_R$ rather than $s^2_R(\bm x)$ itself, as it contains the same information but has the same dimensionality as the density field.

Although the present work focuses on the numerical performance of different marks, it is useful to understand the perturbative structure of the shear mark in Fourier space. In particular, the quadratic shear invariant $s^2_R(\bm x)$ can be written as a convolution of the smoothed density field with a quadrupolar kernel, and the marked field $\tau=s_R(1+\delta)-\langle s_R(1+\delta)\rangle$ admits then a perturbative expansion around the mean shear value. These expressions, summarised in Appendix \ref{app:fourier}, make explicit that the marked power spectra receive contributions from higher-order correlation functions beyond the ordinary matter power spectrum. They provide the starting point for a future perturbative interpretation of the bispectrum- and trispectrum-like information accessed by shear-based marks, which is beyond the scope of this work.

\item \textbf{The isotropy mark.}
Inspired by the White mark, our first family of marks preferentially upweights regions with nearly isotropic tidal fields. Its functional form is
\begin{equation}\label{eqn:iso_mark}
\mathrm{Iso}(\bm{x};R,p,b)=\left[1+b\,s_R^2(\bm{x})\right]^{-p},
\end{equation}
where the parameter $b$ sets the characteristic shear scale over which the weighting varies, while $p$ controls the strength of the weighting. Since the shear vanishes only for perfectly isotropic configurations, positive values of the exponent $p$ assign larger weights to both cluster-like and void-like environments while suppressing anisotropic structures such as filaments and walls. 

Although this mark preferentially weights isotropic environments, it should not be interpreted as a pure cluster-and-void selector. While the centres of virialised haloes are typically close to isotropic and therefore receive large weights, their surroundings often exhibit both high density and strong tidal anisotropy. Therefore, the isotropy mark suppresses part of the cluster outskirts and is not a perfect isotropic environment tracer.

\item \textbf{The filament mark.}
Filamentary environments are characterised by simultaneously exhibiting significant tidal anisotropy while avoiding the highest-density cluster cores. To isolate these regions, we define the filament mark as
\begin{equation}\label{eqn:fil_mark}
\mathrm{Fil}(\bm{x};R,\delta_\ast)=
s_R^2(\bm{x})
\exp\left[-\frac{\delta_R^2(\bm{x})}{2\delta_\ast^2}\right].
\end{equation}
The shear factor enhances anisotropic regions, whereas the Gaussian suppression in density reduces the contribution from highly overdense environments. The parameter $\delta_\ast$ controls the density scale above which this suppression becomes significant.

The Gaussian suppression in density is included because tidal anisotropy and overdensity are only weakly related in the linear regime. As non-linear gravitational collapse proceeds, however, larger values of the tidal eigenvalues generally produce both larger overdensities and larger values of $s^2$. Without the density-dependent suppression, the shear term alone would therefore increasingly favour the densest cluster environments rather than the intermediate-density filamentary structures that this mark is designed to emphasise.

\item \textbf{The Gaussian shear mark.}
This family of marks is designed to isolate environments with a preferred range of tidal anisotropy. Rather than monotonically weighting the shear, it selects a finite interval of shear values through a Gaussian window,
\begin{equation}\label{eqn:gauss_mark}
\mathrm{G}(\bm{x};R,s_0,s_\ast)=
\exp\left[-\frac{\left(s_R(\bm{x})-s_0\right)^2}
{2s_\ast^2}\right].
\end{equation}
The parameter $s_0$ specifies the target shear value, while $s_\ast$ determines the width of the selected range. This family therefore provides a flexible way of probing which levels of tidal anisotropy carry the largest amount of cosmological information.

\item \textbf{Gaussian Local Fractal Dimension (GLFD) mark.} While the previous marks are based on the tidal tensor and therefore probe the anisotropy of the local gravitational field, the LFD provides a complementary geometric characterisation of the cosmic web based on the scaling of the enclosed mass. Moreover, it has the advantage of being a continuous measure of environmental classification, in contrast to the discrete one obtained by the eigenvalue thresholding algorithm. We therefore explore the constraining power of $D_R^{(G)}(\bm x)$, defined in \Cref{eqn:LFD_def_as_ratio}, by using it as a mark. We choose the Gaussian kernel, and not the tophat kernel, mostly due to its superior numerical stability.

\end{itemize}

\begin{figure}[htbp]
    \centering
    \includegraphics[width=0.9\linewidth]{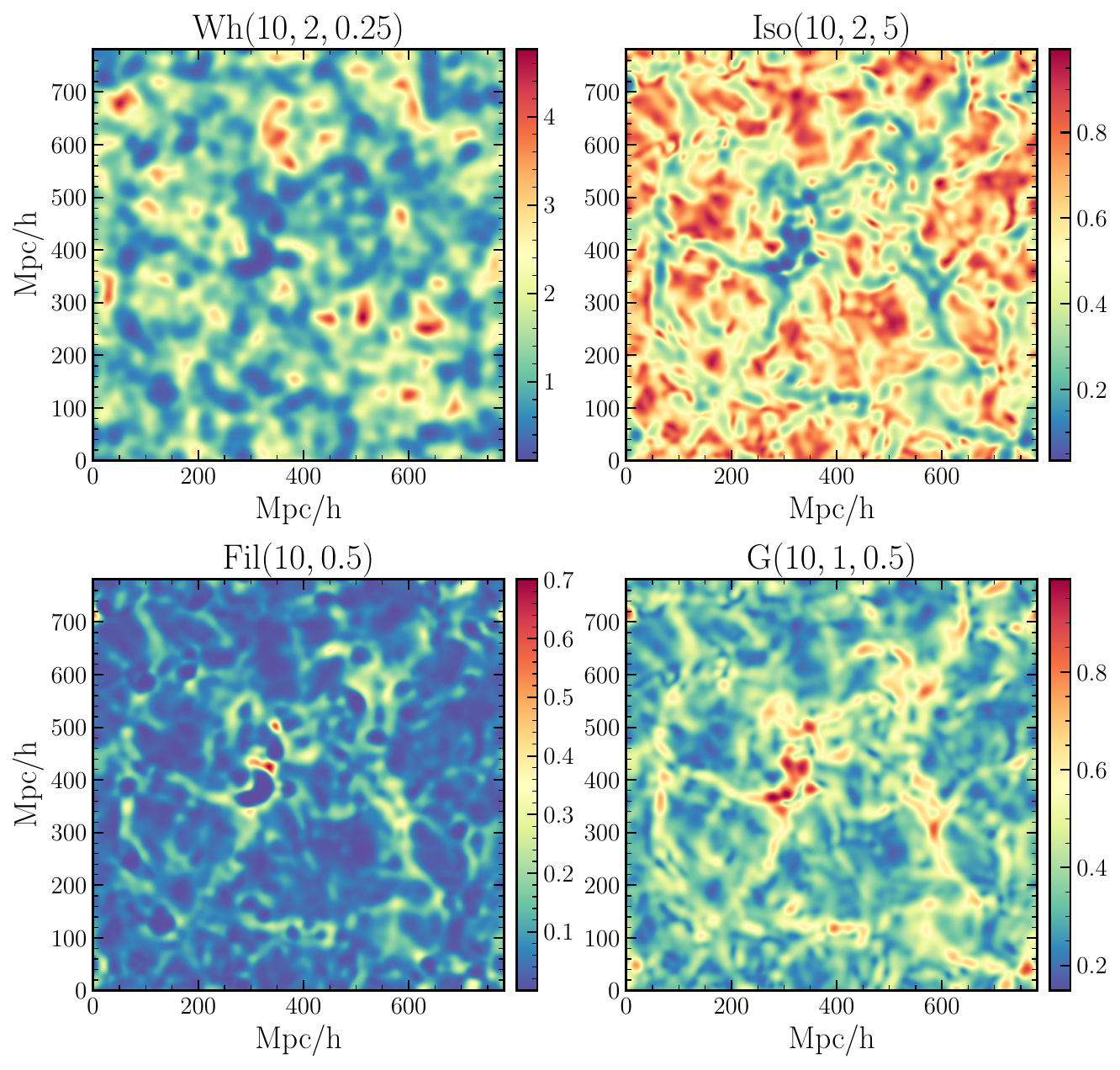}
    \caption{Two-dimensional slices of several mark functions. All fields are smoothed at $R=10$ $h^{-1}$Mpc. The underlying density field is exactly the same as the one in \Cref{fig:delta_and_shear_slices}.}
    \label{fig:elaborated_slices}
\end{figure}

\begin{figure}[htbp]
    \centering
    \includegraphics[width=\linewidth]{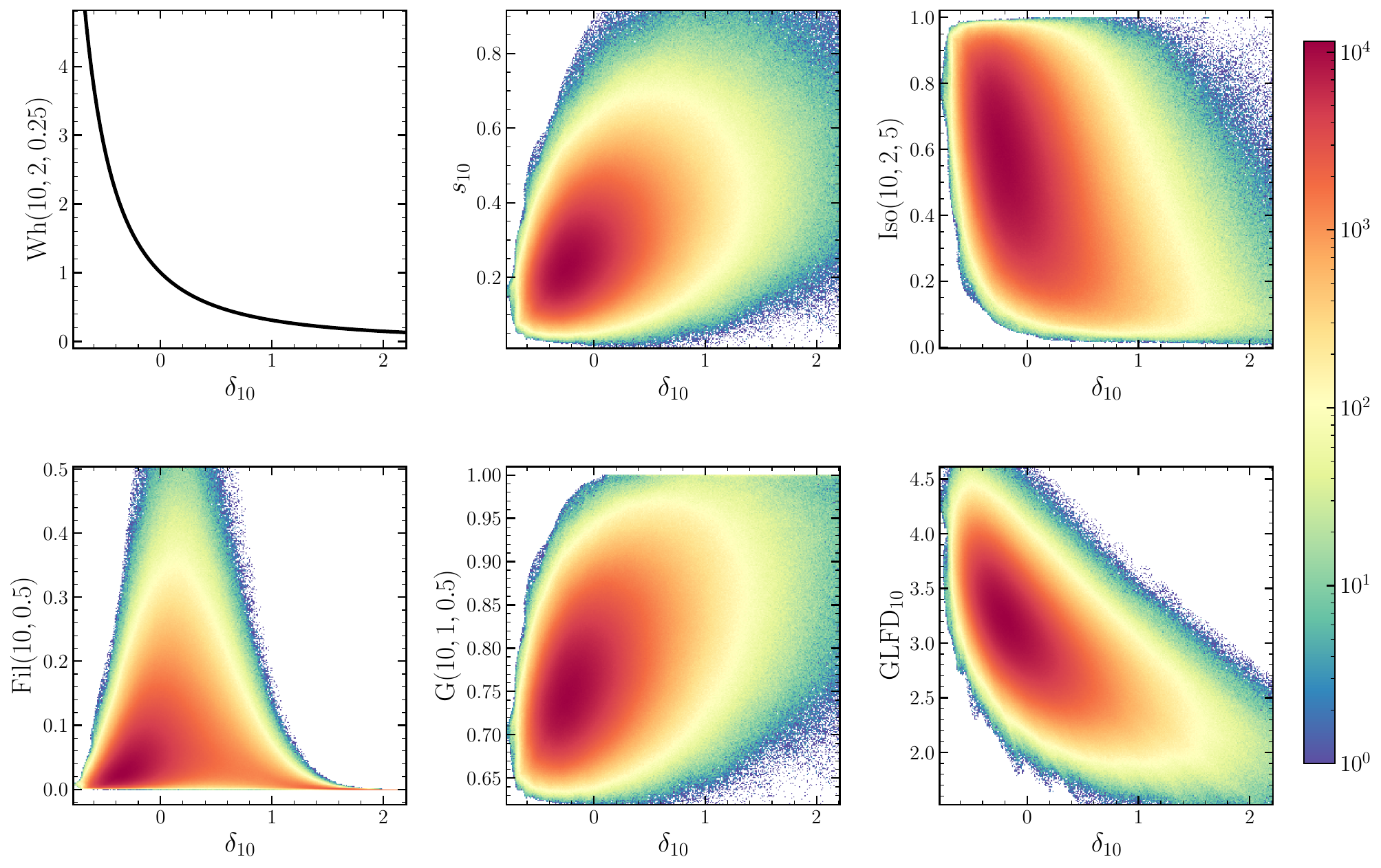}
    \caption{Two dimensional histograms of several marks studied in this work against the underlying smoothed density. Since the White mark is a deterministic function of $\delta_R$, its functional form is plotted rather than a histogram. All fields are smoothed at $R=10$ $h^{-1}$Mpc.}
    \label{fig:2D_hists}
\end{figure}

Example slices of some of these marks are shown in \Cref{fig:elaborated_slices}, while the 2D histograms in \Cref{fig:2D_hists} illustrate their dependence on the underlying density field. Together, these families span a broad range of physically motivated weighting schemes, allowing us to assess which aspects of the local tidal environment provide the greatest improvement over conventional density-based marked statistics. In the remainder of this work we investigate whether these marks capture cosmological information beyond conventional density-based two-point statistics, and quantify their constraining power using suites of $N$-body simulations and Fisher forecasts.

\section{Methodology}\label{sec:methodology}

\subsection{Simulations}\label{ssec:simulations}
The Quijote simulation suite \cite{Villaescusa-Navarro:2019bje} is a publicly available set of more than 43,000 full $N$-body simulations and their derived data products, run for a wide range of cosmological parameters. The subset of simulations used in this work contains $512^3$ cold dark matter particles each (plus an equal number of neutrino particles in the massive-neutrino cosmologies), evolved in a periodic box of size $L_{\text{box}}=1\,h^{-1}\text{Gpc}$ down to redshift $z=0$. We convert the particle snapshots into density fields on a regular grid of resolution $400^3$ using the Cloud-in-Cell (CIC) mass-assignment scheme, as implemented in \textsc{Pylians} \cite{Pylians}. Throughout this work we use only the cold-dark-matter-plus-baryons ($cb$) density field, even in the massive-neutrino cosmologies, since this is the field most closely related to the distribution of galaxies. The fiducial cosmological parameters of the Quijote suite are
\begin{equation}\label{eqn:fid_params}
    \Omega_m=0.3175,\quad\Omega_b=0.049,\quad h=0.6711,\quad n_s=0.9624,\quad \sigma_8=0.834,\quad M_\nu=0.0\text{ eV}\,.
\end{equation}
In addition to the fiducial cosmology, we use the sets of Quijote simulations in which a single parameter is displaced from its fiducial value, namely $\Omega_m^{+/-}=0.3175\pm 0.01$, $\sigma_8^{+/-}=0.834\pm 0.015$, and $M_\nu^{+}=0.1$ eV, $M_\nu^{++}=0.2$ eV, $M_\nu^{+++}=0.4$ eV. These non-fiducial simulations are used to estimate the numerical derivatives of the data vector with respect to the cosmological parameters, as described in the next section.

\subsection{Fisher formalism}\label{ssec:fisher}
We use the Fisher matrix formalism to forecast the constraining power of the marked power spectra on the cosmological parameters of interest. For a single marked field $\Delta$, the corresponding data vector is
\begin{equation}\label{eqn:data_vector_1mark}
    \bm{d}=(P_{\delta\delta}(\bm{k}),P_{\delta\Delta}(\bm{k}),P_{\Delta\Delta}(\bm{k}))^T\,,
\end{equation}
containing the auto-power spectrum of the unmarked field, its cross-power spectrum with the marked field, and the auto-power spectrum of the marked field. When two marked fields, $\Delta$ and $\tau$, are used simultaneously, the data vector is extended to include all of the relevant auto- and cross-power spectra:
\begin{equation}\label{eqn:data_vector_2marks}
    \bm{d}=(P_{\delta\delta}(\bm{k}),P_{\delta\Delta}(\bm{k}),P_{\Delta\Delta}(\bm{k}),P_{\delta\tau}(\bm{k}),P_{\Delta\tau}(\bm{k}),P_{\tau\tau}(\bm{k}))^T\,.
\end{equation}

\begin{figure}[htbp]
    \centering
    \includegraphics[width=0.8\linewidth]{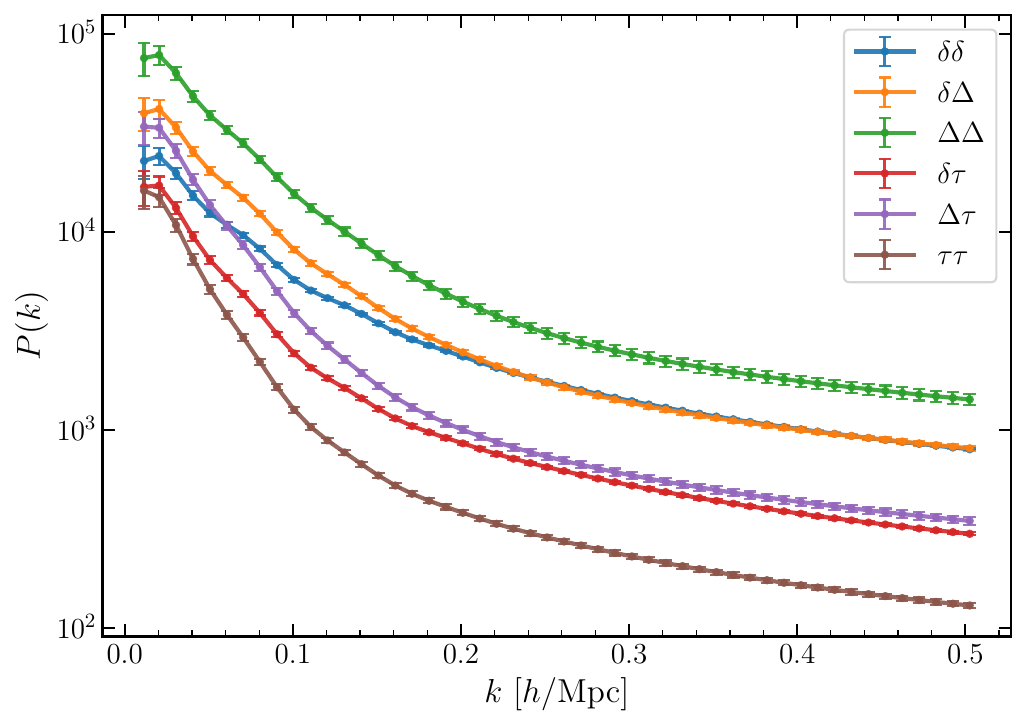}
    \caption{Auto- and cross-spectra of the marked fields $\Delta$ and $\tau$, associated with the $\delta_{10}$ and $s_{10}$ marks respectively, computed as the mean across $N_s=2100$ fiducial realisations. Empirical $1\sigma$ error bars are also plotted.}
    \label{fig:powerspectra_d10s10} 
\end{figure}

All power spectra are evaluated at uniform width bins in the range $[0,k_{\mathrm{max}}]$, where $k_{\mathrm{max}}$ is well below the Nyquist frequency. For the most part of this work we set $k_{\mathrm{max}}=0.3\,h/\mathrm{Mpc}$ and $N_{\mathrm{bins}}=35$, although we also extend part of the analysis to $k_{\mathrm{max}}=0.5\,h/\mathrm{Mpc}$, $N_{\mathrm{bins}}=50$, in order to properly study the dependence of some of our results on $k_{\mathrm{max}}$. As an example, the set of power-spectra associated to the $\delta_{10}$ and $s_{10}$ marks is shown in \Cref{fig:powerspectra_d10s10}.

Assuming that $\bm{d}$ follows a multivariate Gaussian distribution, the Fisher matrix is given by
\begin{equation}\label{eqn:fisher_mat}
    \mathcal{F}_{ij}=\frac{\partial  \bm{d}^T}{\partial\theta_i}\, \mathbf{C}^{-1}\frac{\partial  \bm{d}}{\partial\theta_j}\,,
\end{equation}
where $\mathbf{C}$ is the covariance matrix of the random vector $\bm{d}$, estimated as described in \Cref{ssec:cov_matrices}.

For the evaluation of the Fisher matrix, the covariance matrices $\mathbf{C}$ are never actually inverted directly. Instead, the quadratic form in \Cref{eqn:fisher_mat} is evaluated by means of the Cholesky decomposition of $\mathbf{C}$. Several other works have instead argued that the proper way of computing the Fisher matrix is by using a pseudoinverse approach, obtained from the singular value decomposition of $\mathbf{C}$ after discarding all eigenvalues below a given fraction $f_{\rm thr}$ of the largest one \cite{Park:2022hzj,Cowell:2024wyl}. This last approach has the problem that the dynamic range of the eigenvalues of the covariance matrix strongly depends on the dynamic range of the power spectra involved, and so setting a value for $f_{\rm thr}$ \textit{a priori} becomes difficult. We have checked that the Cholesky approach yields the same results as the pseudoinverse approach as we increase the value of $f_{\rm thr}$, as our covariance matrices are sufficiently well-conditioned.

The cosmological parameters that we choose to vary are
\begin{equation}\label{eqn:fisher_cosmo_params}
    \bm{\theta}=(\Omega_m,\sigma_8,M_\nu)\,.
\end{equation}
This choice is motivated by the fact that $\Omega_m$ and $\sigma_8$ are the parameters most commonly targeted when quantifying the constraining power of mark functions in the literature \cite{Cowell:2024wyl}, while $M_\nu$ is included because certain marked statistics have repeatedly been shown to be especially sensitive to the imprint of neutrino free-streaming on the growth of structure \cite{Massara_2023}, making it a natural parameter against which to test the constraining power of morphological marks. 

\begin{figure}[htbp]
    \centering
    \includegraphics[width=0.95\linewidth]{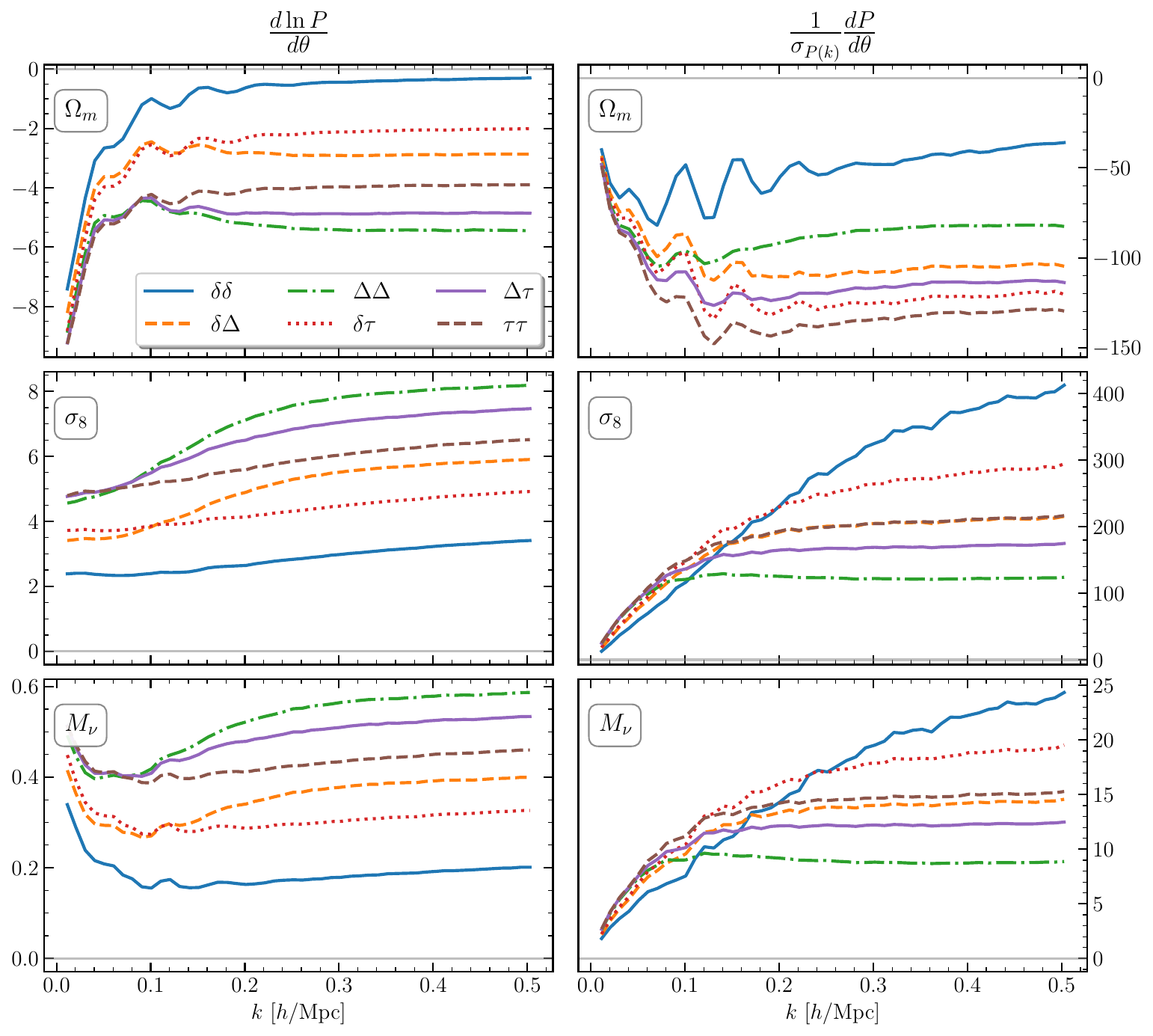}
    \caption{Dependence of the six power spectra of the marked fields $\Delta$ and $\tau$, associated with the $\delta_{10}$ and $s_{10}$ marks respectively, on the cosmological parameters $\Omega_m$, $\sigma_8$, and $M_\nu$. \textit{Left column:} Logarithmic derivatives of the power spectra with respect to each cosmological parameter, $P_{XY}(k)^{-1}\,\partial P_{XY}(k)/\partial\theta$, where $XY\in\{\delta\delta,\delta\Delta,\Delta\Delta,\delta\tau,\Delta\tau,\tau\tau\}$. \textit{Right column:} Corresponding Fisher-information integrands, $\sigma_{P_{XY}}^{-1}(k)\,\partial P_{XY}(k)/\partial\theta$, which quantify the sensitivity of each observable relative to its statistical uncertainty. Larger absolute values indicate greater constraining power per Fourier mode.}
    \label{fig:sensitivity}
\end{figure}

The dependence of the six power spectra associated with the $\delta_{10}$ and $s_{10}$ marks on the cosmological parameters is illustrated in \Cref{fig:sensitivity}, which shows both the logarithmic derivatives (left column) and the corresponding Fisher integrands (right column). It must be noted that the effect of $\sigma_8$ and $M_\nu$ in the power spectra is quite similar, as they both control the amplitudes of $P_{XY}(k)$. The main difference between them is that changing $\sigma_8$ affects the overall amplitude, regardless of scale, whereas the amplitude modulation due to neutrino free-streaming is scale dependent. However, for the range of scales probed in this work, the impact of both parameters seems to be fairly similar, as evident from \Cref{fig:sensitivity}. By computing morphology-dependent marked power spectra in these different cosmologies, we can test whether they respond differently to changes in $\sigma_8$ and $M_\nu$, thereby helping to break the degeneracies that remain when only standard two-point statistics are considered.

The derivatives of the data vector with respect to $\Omega_m$ and $\sigma_8$ are computed with first-order finite differences, using 50 realisations of each non-fiducial cosmology to estimate the corresponding mean data vector,
\begin{equation}\label{eqn:derivative_estimator}
    \frac{\partial  \bm{d}}{\partial\theta_i}\approx \frac{\bm{d}(\theta_i^{fid}+\Delta\theta_i)-\bm{d}(\theta_i^{fid}-\Delta\theta_i)}{2\Delta\theta_i}\,.
\end{equation}
In the case of massive-neutrino cosmologies, however, a different approach must be taken, since the fiducial value $M_\nu=0$ cannot be pushed to more negative values, and \Cref{eqn:derivative_estimator} cannot be applied directly. We therefore follow the same approach as Ref. \cite{Massara_2023} and estimate the derivative with respect to $M_\nu$ by fitting a cubic polynomial through the values $M_\nu=0,0.1,0.2,0.4$ eV, again using 50 realisations per cosmology, and evaluating the derivative of this polynomial at the limiting case $M_\nu=0$:
\begin{equation}\label{eqn:neutrino_derivative_estimator}
    \frac{\partial  \bm{d}}{\partial M_\nu}\approx \frac{5}{6}\left(-21\bm{d}^{\,0}+32\bm{d}^+-12\bm{d}^{++}+\bm{d}^{+++}\right)\,.
\end{equation}
Moreover, extra care must be taken with the massive-neutrino cosmologies, since in the Quijote suite their initial conditions (ICs) are generated at $z=127$ using the Zeldovich approximation, rather than the second-order Lagrangian perturbation theory (2LPT) used for the fiducial cosmology and for the $\Omega_m^{+/-}$ and $\sigma_8^{+/-}$ simulations \cite{Villaescusa-Navarro:2019bje}. We instead evaluate $\bm{d}^{\,0}$ using a dedicated set of 50 fiducial-cosmology realisations generated with Zeldovich ICs, matching those used for the $M_\nu^{+}$, $M_\nu^{++}$, and $M_\nu^{+++}$ simulations \cite{Villaescusa-Navarro:2019bje,Massara_2023}. In \Cref{fig:sensitivity} the derivatives of the $\delta_{10}+s_{10}$ power spectra are shown, normalised both by the power-spectra themselves (left column) and standard deviation of the power-spectra (right column).

\subsubsection{Error Improvement Factors}\label{sssec:eif}
The Fisher matrix provides an estimate of the smallest statistical uncertainty with which each cosmological parameter can be constrained under the assumption of a Gaussian likelihood. We note that this interpretation only strictly holds for Gaussian posteriors; we use it here as a standard first approximation for the level of improvement in parameter constraints, although the exact values may change in a realistic setting. Under this assumption, and marginalizing over all remaining parameters, the forecasted $1\sigma$ uncertainty on $\theta_i$ is given by
\begin{equation}
    \sigma(\theta_i)=\sqrt{(\mathcal{F}^{-1})_{ii}}.
\end{equation}
Since our primary goal is to quantify the amount of cosmological information added by morphological marked fields beyond that already contained in the standard matter power spectrum, it is convenient to express our results in terms of relative improvements rather than absolute parameter uncertainties. To this end, we define the \emph{error improvement factor} (EIF) for a parameter $\theta_i$ as
\begin{equation}\label{eqn:EIF_def}
\mathrm{EIF}(\theta_i)=
\frac{\sigma(\theta_i\,|\,P_{\delta\delta})}
{\sigma(\theta_i\,|\,\bm d)}\,.
\end{equation}
An error improvement factor larger than unity therefore indicates that the marked statistics provide additional cosmological information beyond the conventional power spectrum. For example, $\mathrm{EIF}=2$ implies that the inclusion of the marked observables reduces the forecasted uncertainty on the corresponding parameter by a factor of two. Throughout this work we use the EIF as our primary figure of merit, as it enables a straightforward comparison between different families of marks, smoothing scales, and combinations of observables while remaining independent of the absolute size of the parameter uncertainties.

\subsection{Covariance Matrices}\label{ssec:cov_matrices}
\begin{figure}[htbp]
    \centering
    \includegraphics[width=0.7\linewidth]{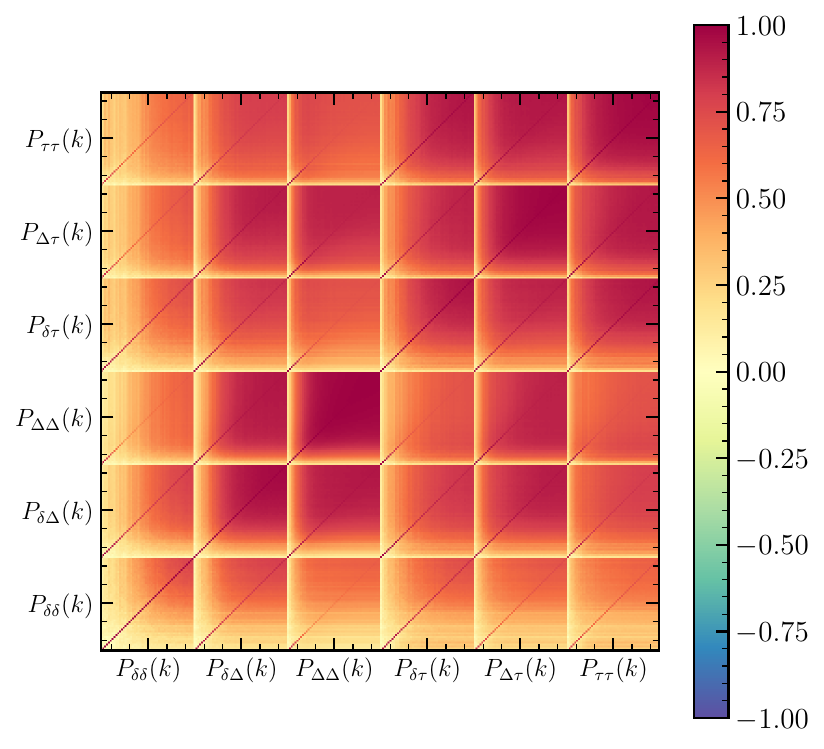}
    \caption{Numerical correlation matrix of the auto- and cross-spectra of the marked fields $\Delta$ and $\tau$, associated with the $\delta_{10}$ and $s_{10}$ marks respectively, estimated from $N_s=2100$ realisations. The maximum wavenumber used for this visualisation is $k_{\rm max}=0.5\,h^{-1}\text{Mpc}$ }
    \label{fig:corr_d10s10}
\end{figure}

We estimate the covariance matrix $\mathbf{C}$ of the data vector $\bm{d}$ from $N_s=2,100$ realisations\footnote{This number of simulations was chosen because the largest data vector considered in this work contains $N_d=6\times 35=210$ elements, so that $N_s\approx 10\times N_d$. This is a commonly adopted rule of thumb that ensures a stable estimate of the off-diagonal elements of the covariance matrix.} of the fiducial cosmology, using the standard unbiased sample covariance estimator 
\begin{equation}\label{eqn:cov_estimator}
    \widehat{\mathbf{C}} = \frac{1}{N_s-1}\sum_{i=1}^{N_s}\left(\bm{d}_i-\bar{\bm{d}}\right)\left(\bm{d}_i-\bar{\bm{d}}\right)^T\,,
\end{equation}
where $\bm{d}_i$ is the data vector measured in the $i$-th fiducial realisation and $\bar{\bm{d}}$ is the corresponding sample mean.
Because the covariance matrix is inferred from a finite number of independent mock realisations, its inverse (the precision matrix) is a biased estimate of the true inverse covariance. To correct for this and obtain an unbiased estimator, we apply the Hartlap factor \cite{Hartlap:2006kj} whenever the inverse covariance matrix is used:
\begin{equation}\label{eqn:hartlap_factor}
\widehat{\mathbf{C}^{-1}} = \frac{N_s - N_d - 2}{N_s - 1} \,\widehat{\mathbf{C}}^{-1}\,,
\end{equation}
where $N_d$ is the total length of the data vector. Finally, we have assessed the convergence of our results by checking that the derived parameter constraints are stable under an increase in the number of realisations used for the covariance estimate. \Cref{fig:corr_d10s10} shows the correlation matrix of the $\delta_{10}$ and $s_{10}$ marked power spectra.

A quick inspection of \Cref{fig:corr_d10s10} reveals that this particular covariance matrix is far from diagonal. Significant correlations are present not only between neighbouring wavenumbers, but also between different spectra. This observation prompts two important comments:

\begin{enumerate}

\item On the one hand, it goes to show that approximating the covariance matrix with the Gaussian covariance approximation
\begin{equation}
    \mathrm{Cov}(P_{AB}(k),P_{CD}(k'))=\frac{\delta^{\rm K}_{kk'}}{N_k}[P_{AC}(k)\,P_{BD}(k)+P_{AD}(k)\,P_{BC}(k)]\,,
\end{equation}
where $N_k$ is the number of modes used to calculate the power spectra in the corresponding bin, is almost certainly not adequate. To assess how large the beyond-Gaussian correlations were, we compared both the diagonal elements of the numerical and Gaussian covariance matrices. An example is shown in \Cref{fig:num_vs_gauss_cov}. While the Gaussian prediction reproduces the covariance well on the largest scales, it increasingly underestimates and misses the strong mode coupling towards larger wavenumbers. Using the Gaussian approximation can thus lead to overestimating the Fisher constraints, so for EIF computations we use numerical covariances for which we have ensured convergence with the number of realisations.

\begin{figure}[htbp]
    \centering
    \includegraphics[width=\linewidth]{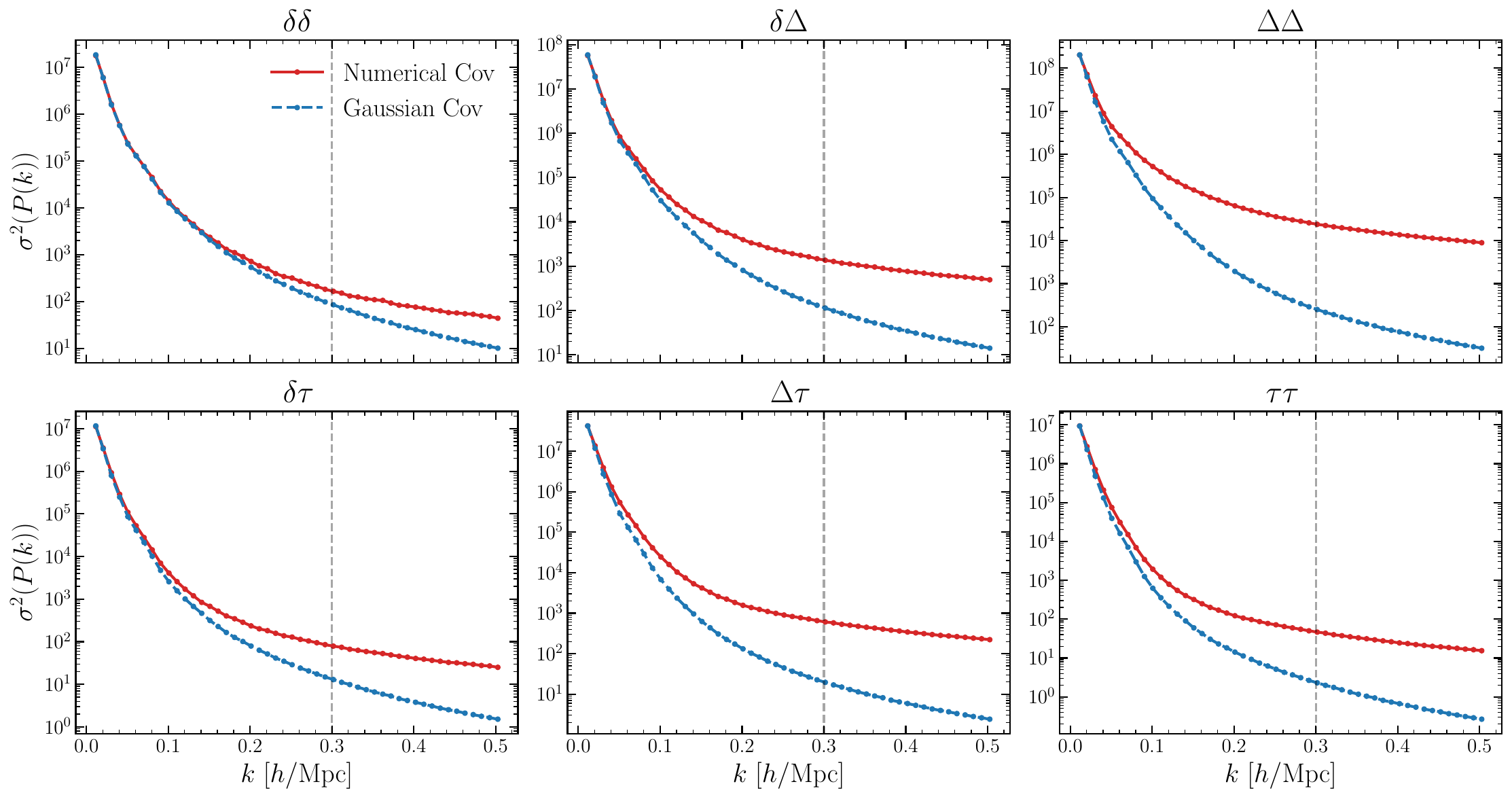}
    \caption{Comparison between the diagonal elements of the numerical covariance matrix (solid red) and the Gaussian covariance approximation (blue dashed) for the six auto- and cross-spectra of the marked fields $\Delta$ and $\tau$, associated with the $\delta_{10}$ and $s_{10}$ marks respectively. The vertical dashed lines show the fiducial value of $k_{\rm max}=0.3\,h^{-1}\text{Mpc}$ adopted throughout this work.}
    \label{fig:num_vs_gauss_cov}
\end{figure}

\item On the other hand, it must be emphasised that the presence or absence of off-diagonal correlations should not be interpreted as a direct indicator of constraining power of a marked statistic. Since the Fisher information is invariant under invertible affine transformations of the marks, seemingly different covariance matrices can yield identical EIFs: a more diagonal covariance matrix is not intrinsically more informative. As a concrete example, consider the White mark $m(\bm{x})$ and the transformed mark $\tilde m(\bm{x})=1-m(\bm{x})$. The covariance matrices of the data vectors associated with these marks are related to each other via a simple linear transformation that mixes all rows and columns. \Cref{fig:white_vs_1mwhite} illustrates this point. The correlation matrix associated with the original White mark exhibits almost no correlation among different wavenumbers, whereas the equivalent matrix for the transformed mark exhibits substantial off-diagonal structure. Nevertheless, both marks yield identical Fisher constraints and therefore identical EIFs. This simple example demonstrates that the apparent ``cleanliness'' of a covariance matrix should not be interpreted as evidence that a statistic is intrinsically more informative. The superior performance of the White mark must instead originate from the cosmological information encoded in the marked field itself, rather than from the visual appearance of its covariance matrix.

\begin{figure}[htbp]
    \centering
    \includegraphics[width=\linewidth]{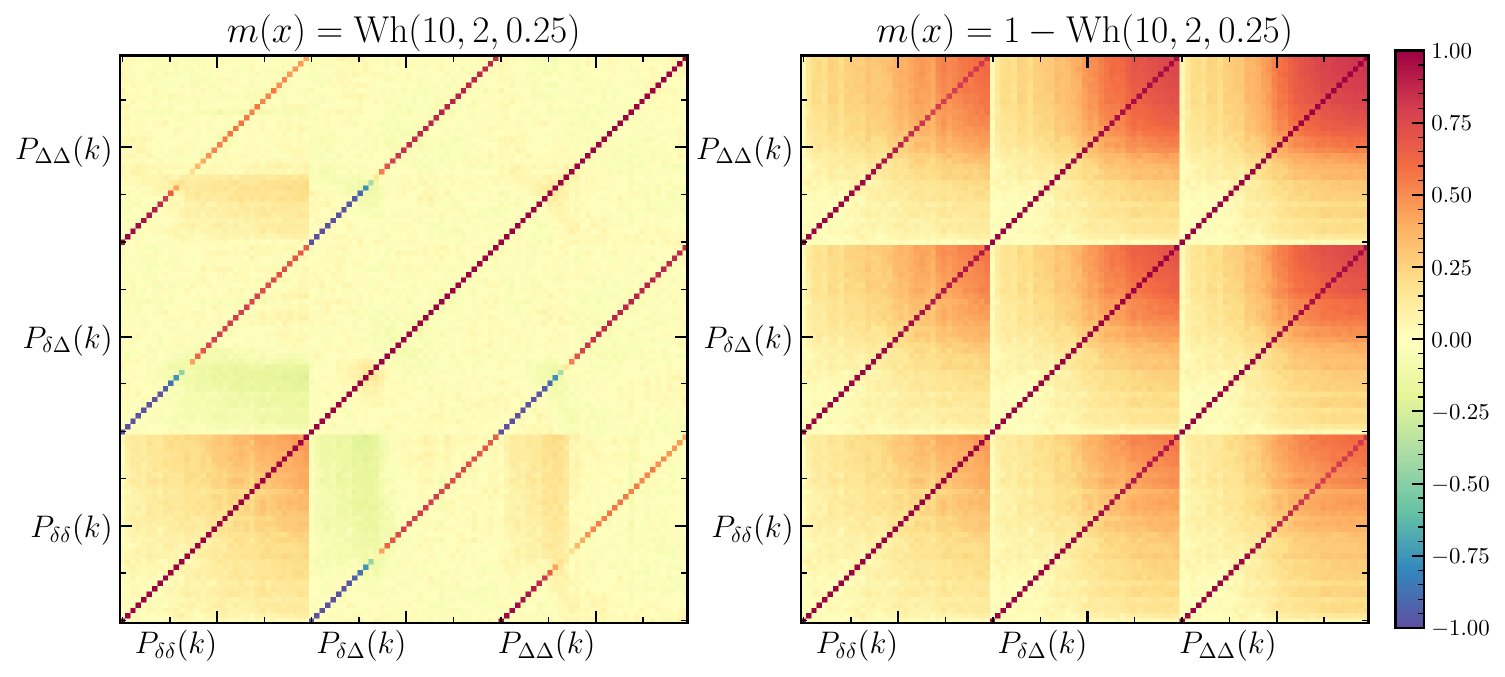}
    \caption{Correlation matrices of the data vectors associated with the White mark (left) and one minus the White mark (right). Both matrices are related to each other via a congruence transformation.}
    \label{fig:white_vs_1mwhite}
\end{figure}
\end{enumerate}

\subsection{Smoothing Scale}\label{ssec:smoothing_scale}

The choice of smoothing scale is very important when designing marked statistics, as it can control the type of information that is extracted. As demonstrated in \cite{Cowell:2025mov}, applying a mark smoothed on very small scales can inject Gaussian variance from highly non-linear, very small scale modes into the marked two-point statistics, artificially inflating the apparent constraining power without extracting genuinely non-Gaussian information.\footnote{This effect was demonstrated in Ref.~\cite{Cowell:2025mov} for the autocorrelation of density-marked fields. Its relative importance depends on the particular definition of the mark, although the underlying mechanism is expected to be more general.} The word ``artificially'' here refers to the fact that this information could have been recovered by simply extending the usual density power spectrum to larger wavenumbers. Because our primary goal is to isolate and constrain the non-Gaussian morphological information to which the standard two-point function is blind, we focus on smoothing scales ($R \geq 5 \, h^{-1}\text{Mpc}$) where the resulting error improvement is dominated by the reorganization of mode-coupling and non-Gaussianity, rather than the leakage of small-scale Gaussian variance.

We choose as our fiducial smoothing scale $R = 10 \, h^{-1}\text{Mpc}$, which represents a compromise between resolving the large-scale web and avoiding an excessive sensitivity to highly non-linear small-scale fluctuations. We nevertheless partially explore the dependence of the EIFs on the smoothing scale in \Cref{ssec:results_part1}.

\section{Results and Discussion}\label{sec:results}
In this section we report the EIF associated to the morphological marks described in \Cref{sssec:morpho_marks}. First, we analyse the results of using the shear amplitude field $s_R(\bm x)$ as a mark in \Cref{ssec:results_part1}. Next, in \Cref{ssec:results_part2} we report our findings for the more physically-motivated isotropy, filament and Gaussian shear marks. Finally, in \Cref{ssec:results_part3} we show the performance of the Gaussian LFD mark.

\subsection{Shear as a mark}\label{ssec:results_part1}

Before focusing on more elaborate morphology-inspired marks, we first investigate whether the shear amplitude itself as a mark can surface useful cosmological information. Since both $\delta_R$ and $s_R$ are low-order tidal invariants with relatively simple perturbative descriptions, they also provide a useful benchmark for understanding how morphology-dependent information enters marked statistics.

\begin{figure}[htbp]
    \centering
    \includegraphics[width=\linewidth]{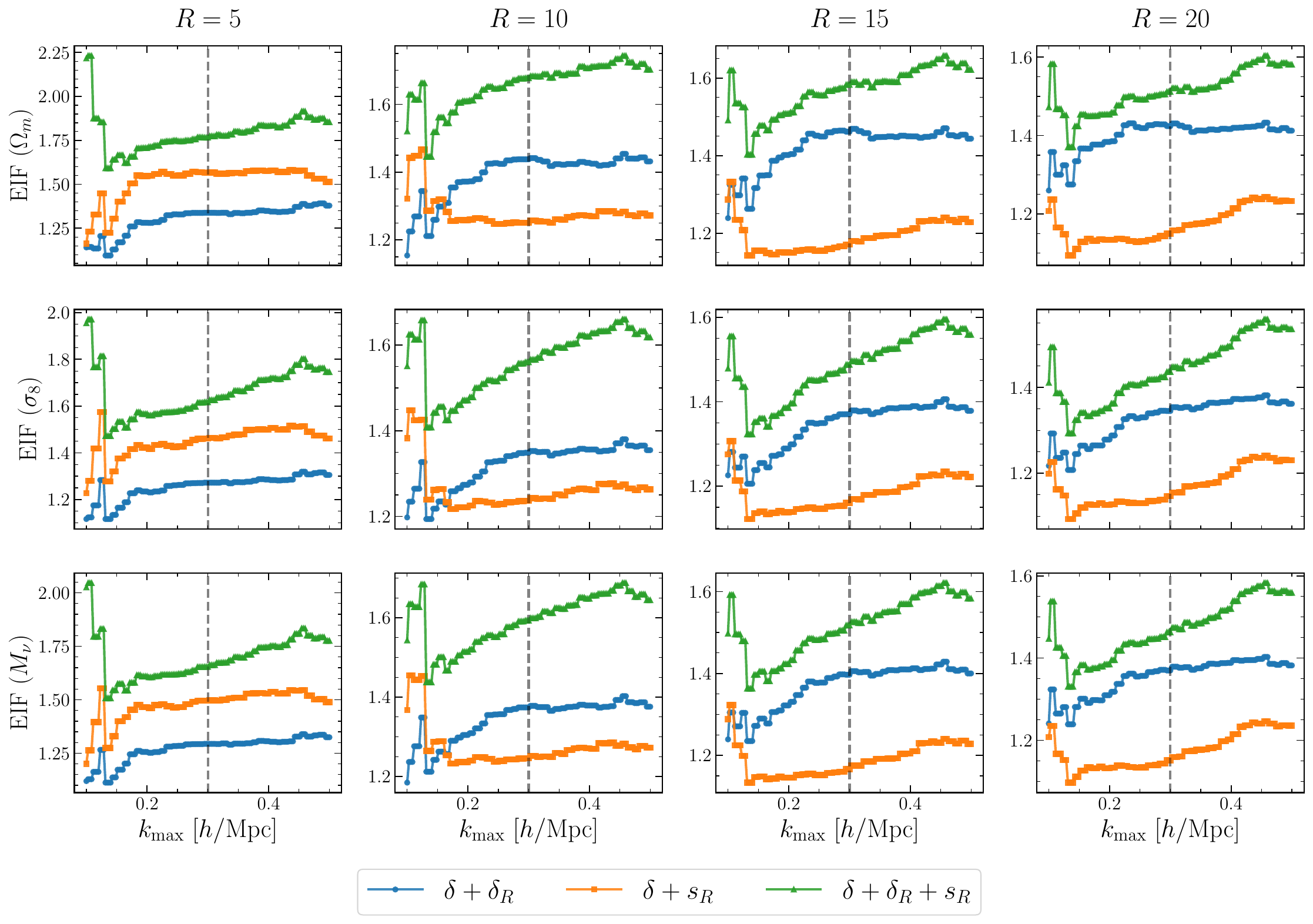}
    \caption{Error Improvement Factors associated to the smoothed density mark (blue lines), smoothed shear mark (orange lines) and the combination of smoothed density and shear marks (green lines) as a function of the maximum value of wavenumber $k$ included in the Fisher analysis. Results are shown for 4 different values of the smoothing scale $R$. The vertical dashed lines show the fiducial value of $k_{\rm max}=0.3\,h^{-1}\text{Mpc}$ adopted throughout this work.}
    \label{fig:EIF_vs_kmax}
\end{figure}

The resulting EIFs are summarised in \Cref{tab:eif_results_table_part1}, while their dependence on the maximum wavenumber included in the Fisher analysis is shown in \Cref{fig:EIF_vs_kmax}. We compare three different data vectors: the standard density field combined with the smoothed density mark ($\delta+\delta_R$), the density field combined with the shear mark ($\delta+s_R$), and the joint analysis including both marks ($\delta+\delta_R+s_R$). We also vary the smoothing scale $R$ in the range $[5\,h^{-1}{\rm Mpc},\,20\,h^{-1}{\rm Mpc}]$.

\begin{table}[htpb]
    \centering

    \begin{tabular}{ccccc}
        \toprule
        \multirow{2}{*}{\textbf{Configuration}} & \multirow{2}{*}{$\boldsymbol{R \ [\text{Mpc/h}]}$} & \multicolumn{3}{c}{\textbf{EIF}} \\
        \cmidrule(lr){3-5}
        & & $\boldsymbol{\Omega_m}$ & $\boldsymbol{\sigma_8}$ & $\boldsymbol{M_\nu}$ \\
        \midrule
        \multirow{4}{*}{$\delta + \delta_R$} 
        & $5$   & $1.38$ & $1.31$ & $1.33$ \\
        & $10$  & $1.43$ & $1.35$ & $1.38$ \\
        & $15$  & $1.44$ & $1.38$ & $1.40$ \\
        & $20$  & $1.42$ & $1.36$ & $1.38$ \\
        \midrule
        \multirow{4}{*}{$\delta + s_R$} 
        & $5$   & $1.52$ & $1.47$ & $1.50$ \\
        & $10$  & $1.27$ & $1.26$ & $1.27$ \\
        & $15$  & $1.23$ & $1.23$ & $1.23$ \\
        & $20$  & $1.24$ & $1.24$ & $1.24$ \\
        \midrule   
        \multirow{4}{*}{$\delta + \delta_R + s_R$} 
        & $5$   & $1.87$ & $1.76$ & $1.79$ \\
        & $10$  & $1.70$ & $1.62$ & $1.64$ \\
        & $15$  & $1.63$ & $1.57$ & $1.59$ \\
        & $20$  & $1.59$ & $1.55$ & $1.57$ \\
        \bottomrule
            
    \end{tabular}
    \caption{Numerical Error Improvement Factors (EIF) associated with the smoothed density mark $(\delta + \delta_R)$, shear amplitude mark $(\delta + s_R)$ and combination of both marks ($\delta + \delta_R + s_R$) for different representative smoothing scales.}
    \label{tab:eif_results_table_part1}
\end{table}

The first conclusion is that the shear amplitude field by itself contains a rather modest amount of constraining power. For the fiducial smoothing scale $R=10\,h^{-1}{\rm Mpc}$, using $s_R$ produces EIFs of approximately $1.27$ for all three cosmological parameters. In comparison, the conventional smoothed density mark consistently performs better, reaching EIFs of $1.35-1.43$ for the same smoothing scale. The natural question then is: how complementary is this information that each mark is bringing? When both marks are combined, across all smoothing scales considered, the joint data vector $(\delta+\delta_R+s_R)$ substantially outperforms either individual mark, yielding improvements of $60-90\%$ over the density power spectrum alone. This seems to demonstrate that the shear field contains information that is largely complementary to that encoded in the smoothed density.

The dependence on the smoothing scale reveals an interesting difference between the density and shear marks. Whereas the performance of the density mark is relatively insensitive to $R$, the constraining power of the shear mark decreases noticeably as the smoothing scale increases. The strongest improvements are obtained for $R=5\,h^{-1}{\rm Mpc}$, after which the EIF gradually declines. This behaviour is consistent with the interpretation of the shear field as a tracer of anisotropic structures, whose contrast is progressively ``washed out'' by larger smoothing scales.

\Cref{fig:EIF_vs_kmax} shows how the EIFs vary with the maximum wavenumber $k_{\rm max}$ included in the analysis. As $k_{\rm max}$ increases, new Fourier modes are included in the Fisher analysis, but the binning in $k$ is kept fixed. Therefore, increasing $k_{\rm max}$ always increases the total amount of information available, and the marginalized uncertainties $\sigma(\theta)$ on the cosmological parameters decrease monotonically for every data vector considered. The EIF, however, measures the \emph{ratio} between the parameter uncertainties obtained with and without the marked statistics. Therefore, its value need not increase monotonically with $k_{\rm max}$. A decrease in the EIF does not imply that information has been lost; rather, it indicates that the standard density power spectrum is gaining information faster over that range of scales than the corresponding marked data vectors. Similarly, an increasing EIF means that the marked observables are extracting information from the newly added modes more efficiently than the density field alone.

\begin{figure}[htbp]
    \centering
    \includegraphics[width=\linewidth]{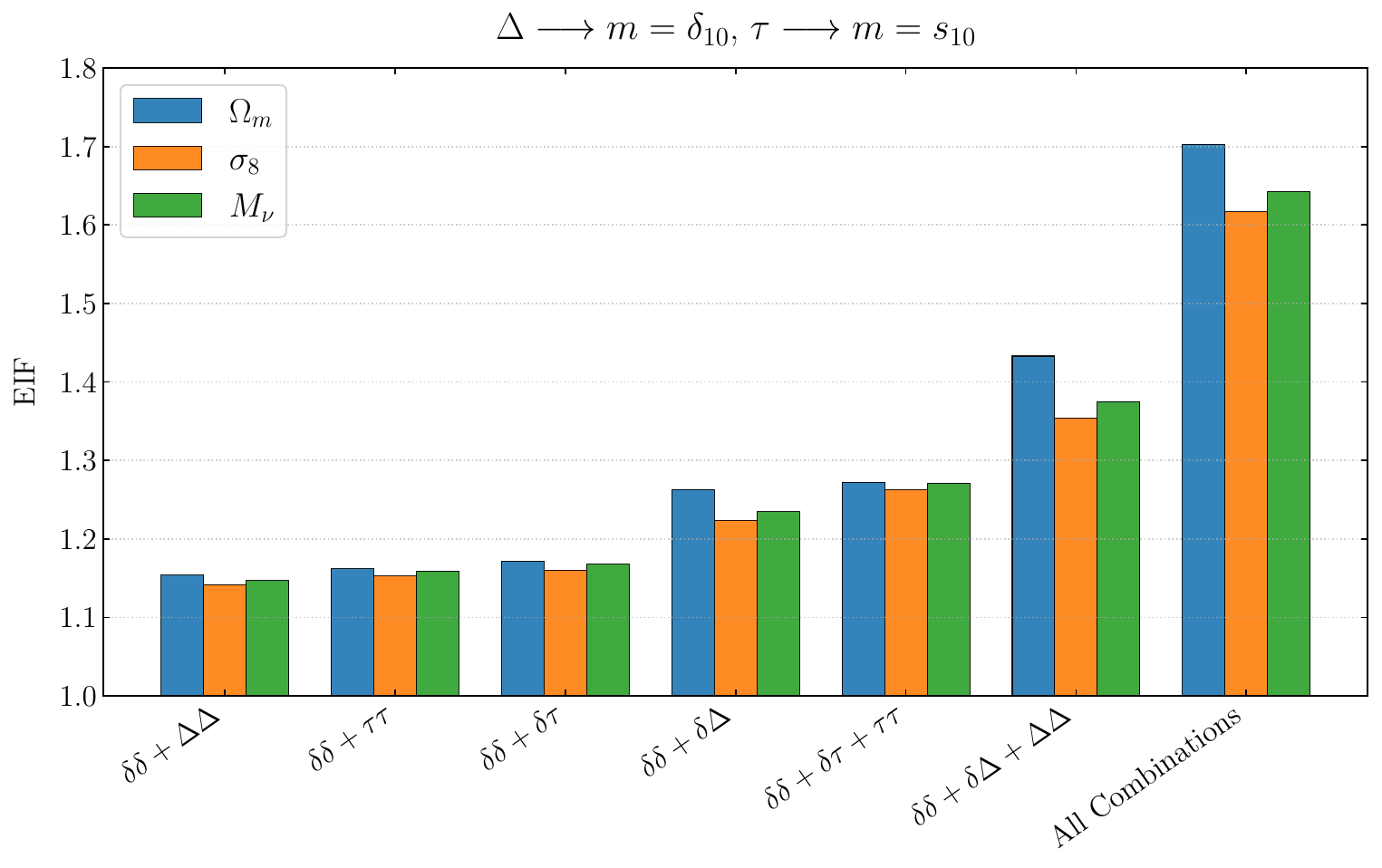}
    \caption{Error Improvement Factors corresponding to different combinations of power spectra of the marked fields $\Delta$ and $\tau$, associated with the $\delta_{10}$ and $s_{10}$ marks respectively.}
    \label{fig:barchart_EIF}
\end{figure}

Finally, \Cref{fig:barchart_EIF} provides insight into the origin of the additional information by decomposing the Fisher analysis into different combinations of power spectra. The cross-spectrum between the density field and the $\delta_R$-marked field contributes substantially more than the auto-spectrum of the marked field. For example, the combination $(P_{\delta\delta},P_{\delta\Delta})$ already captures most of the improvement obtained from the full $(\delta+\delta_R)$ analysis, whereas adding only the marked auto-spectrum produces a comparatively modest gain. This behaviour is consistent with the perturbative expectation of marked statistics: the cross-spectrum receives its leading non-Gaussian contributions from bispectrum-like terms, while marked auto-spectra are more sensitive to trispectrum contributions, which are typically noisier and contain less independent information \cite{Ebina:2024zkv}. In the case of the shear, such a simple interpretation does not directly apply, because the shear amplitude is a non-linear functional of the density field, so both observables receive contributions from multiple higher-order correlations. Although the exact hierarchy of the different power spectrum combinations varies somewhat with smoothing scale, the overall conclusions remain unchanged. The corresponding results for all smoothing scales are shown in Appendix \ref{app:shear_results}.

\subsection{Targeted shear marks}\label{ssec:results_part2}
We now investigate whether the physically motivated shear-based marks introduced in \Cref{sssec:morpho_marks} outperform the simple shear-amplitude mark. Additionally, we are interested in quantifying whether targeted shear marks can yield a significant error improvement when combined with a similarly physically-motivated, density-dependent mark function, such as the White mark. Throughout this section we fix the smoothing scale to the fiducial value $R=10\,h^{-1}{\rm Mpc}$. The results are summarised in \Cref{tab:eif_wh10_combinations}.
\begin{table*}[htpb]
  \centering
  \begin{tabular}{ccccc}
    \toprule
    \multirow{2}{*}{\textbf{Configuration}} & \multirow{2}{*}{$\boldsymbol{\tau}$} & \multicolumn{3}{c}{\textbf{EIF}} \\
    \cmidrule(lr){3-5}
    & & $\boldsymbol{\Omega_m}$ & $\boldsymbol{\sigma_8}$ & $\boldsymbol{M_\nu}$ \\
    \midrule
    $\delta + \text{Wh}_{10}$ & White Mark ($R=10, p=2, b=0.25$) & $2.24$ & $1.87$ & $1.93$ \\
    \midrule
    \multirow{11}{*}{$\delta + \tau$} 
    & $s_{10}$ & 1.27 &  1.26  & 1.27 \\ 
    \addlinespace
    & Gauss ($s_0=1, s_*=0.5$) & 1.26 & 1.24 & 1.25 \\        
    & Gauss ($s_0=1, s_*=1$) & 1.27 & 1.26 & 1.26 \\      
    & Gauss ($s_0=2, s_*=0.5$) & 1.14 & 1.14 & 1.15 \\ 
    & Gauss ($s_0=2, s_*=1$) & 1.23 & 1.22 & 1.23 \\
        
    \addlinespace
    & Fil ($\delta_*=0.50$) & $1.32$ & $1.25$ & $1.26$ \\
    & Fil ($\delta_*=1.00$) & $1.26$ & $1.23$ & $1.24$ \\
    \addlinespace
    & Iso ($b=1.0, p=1.0$) & $1.24$ & $1.23$ & $1.24$ \\
    & Iso ($b=1.0, p=2.0$) & $1.26$ & $1.24$ & $1.25$ \\
    & Iso ($b=5.0, p=1.0$) & $1.28$ & $1.26$ & $1.27$ \\
    & Iso ($b=5.0, p=2.0$) & $1.27$ & $1.26$ & $1.26$ \\
    \midrule
    \multirow{11}{*}{$\delta + \text{Wh}_{10} + \tau$} 
    & $s_{10}$ & 2.46 & 2.05 & 2.13 \\    
    \addlinespace
    & Gauss ($s_0=1, s_*=0.5$) & 2.43 & 2.03 & 2.10 \\    
    & Gauss ($s_0=1, s_*=1$) & 2.44 & 2.03 & 2.10 \\
    & Gauss ($s_0=2, s_*=0.5$) & 2.44 & 2.05 & 2.12 \\
    & Gauss ($s_0=2, s_*=1$) & 2.46 & 2.05 & 2.13 \\
    \addlinespace
    & Fil ($\delta_*=0.50$) & $2.49$ & $2.08$ & $2.15$ \\
    & Fil ($\delta_*=1.00$) & $2.51$ & $2.09$ & $2.17$ \\
    \addlinespace
    & Iso ($b=1.0, p=1.0$) & $2.45$ & $2.05$ & $2.12$ \\
    & Iso ($b=1.0, p=2.0$) & $2.45$ & $2.04$ & $2.11$ \\
    & Iso ($b=5.0, p=1.0$) & $2.45$ & $2.04$ & $2.11$ \\
    & Iso ($b=5.0, p=2.0$) & $2.43$ & $2.03$ & $2.10$ \\
    \bottomrule
  \end{tabular}
  \caption{Numerical Error Improvement Factors (EIF) for a variety of mark combinations. All fields are smoothed at $R=10\,\rm{Mpc}/h$.}
  \label{tab:eif_wh10_combinations}
\end{table*}
The first thing to note is that none of the targeted shear marks beats the performance of the standard White mark when used individually; their EIFs remain in the range $1.2-1.3$, substantially below the White mark, which reaches EIFs of approximately $2.2$, $1.9$ and $1.9$ for $\Omega_m$, $\sigma_8$ and $M_\nu$, respectively. 

A striking feature of \Cref{tab:eif_wh10_combinations} is the weak dependence on the exact functional form of the elaborated shear mark. Whether one enhances isotropic environments, filamentary regions, or a preferred range of shear values, the resulting EIFs differ only at the few-percent level and basically match those obtained by the shear amplitude mark itself. This suggests that the dominant source of cosmological information is the shear field itself, while the precise non-linear transformation used to construct the mark plays only a secondary role.

The more interesting result is obtained when the shear-based marks are combined with the White mark. In every case the joint data vector $(\delta+\mathrm{Wh}_{10}+\tau)$ outperforms the White mark alone, suggesting that the shear marks probe information that is not fully captured by density weighting. Although the improvement is modest, it is consistent across all cosmological parameters and all families of shear marks: the constraints on  all three parameters improve by $\sim 10\%$. These findings suggest that it is possible to complement the White mark with tidal shear based marks to further constrain cosmological parameters.

Finally, we emphasise that this comparison has been carried out only for the fiducial smoothing scale $R=10\,h^{-1}{\rm Mpc}$. Although the shear mark yields slightly larger EIFs at $R=5h^{-1}\text{Mpc}$, we retain $R=10h^{-1}\text{Mpc}$ throughout this section as a more conservative fiducial choice. This value of $R$ lies comfortably above the grid resolution of the density field and reduces the impact of pixelisation and other numerical artifacts, while remaining sufficiently small to resolve the cosmic-web structure. Moreover, as discussed in \Cref{ssec:smoothing_scale} our goal here is not to maximise the constraining power by pushing the smoothing scale into the highly non-linear regime, but rather to isolate the non-Gaussian information encoded in the morphology of the matter distribution. It is therefore plausible that the physically motivated shear marks could achieve somewhat better performance at smaller smoothing scales, but a dedicated optimisation of both their hyperparameters and smoothing scale is beyond the scope of this work.

\subsection{Other morphological marks}\label{ssec:results_part3}

We conclude by exploring marks based on alternative geometric descriptions of the cosmic web. In particular, we consider the Gaussian Local Fractal Dimension (GLFD) mark introduced in \Cref{sssec:morpho_marks}. The performance of this mark is summarised in \Cref{tab:eif_results_part3}. We also experimented with marks based on measures of local non-sphericity, $\mathrm{NS}(\bm x)=\lambda_1-\lambda_3$, but these consistently produced smaller improvements than the shear- and LFD-based constructions and are therefore not discussed further.
\begin{table}[htpb]
    \centering
    \begin{tabular}{cccc}
        \toprule
        \multirow{2}{*}{\textbf{Configuration}} &  \multicolumn{3}{c}{\textbf{EIF}} \\
        \cmidrule(lr){2-4}
        &  $\boldsymbol{\Omega_m}$ & $\boldsymbol{\sigma_8}$ & $\boldsymbol{M_\nu}$ \\
        \midrule
        $\delta$ + Wh$_{10}$& $2.24$ & $1.87$ & $1.93$ \\
       $\delta$ + GLFD$_{10}$& $1.63$ & $1.46$ & $1.51$ \\        
       $\delta + \delta_{10}+\delta_{15}$ & $1.52$ & $1.43$ & $1.46$ \\    
        \midrule
       $\delta$ + Wh$_{10}$+ GLFD$_{10}$ & $2.45$ & $2.07$ & $2.14$ \\
        \bottomrule
            
    \end{tabular}
    \caption{Numerical Error Improvement Factors (EIF) for some given mark combinations.}
    \label{tab:eif_results_part3}
\end{table}
The GLFD mark provides a clear improvement over the standard density power spectrum, yielding EIFs of $1.63$, $1.46$, and $1.51$ for $\Omega_m$, $\sigma_8$, and $M_\nu$, respectively. Although these gains remain below those achieved by the White mark alone, they demonstrate that the LFD constitutes a useful morphology-sensitive summary of the density field.

The two-dimensional histograms presented in \Cref{fig:2D_hists} show that the GLFD exhibits a qualitative dependence on density similar to that of the White mark, assigning its largest values to underdense environments while suppressing dense regions. This naturally suggested that the GLFD could inherit some of the aspects that make the White mark successful. Our results confirm this expectation only partially: although the GLFD is indeed an effective mark, it does not outperform the empirically optimized White mark. It must be noted that the White mark has two free hyperparameters to tune, while the GLFD has none.

Another interesting comparison is provided by multiscale density information. Since the derivative of the enclosed mass can be partially approximated as the difference of the densities at two different smoothing scales, it is natural to ask whether the information gain coming from the GLFD is similar to that obtained by using smoothed fields at two different smoothing scales as marks. The data vector consisting of the density field together with two smoothed density marks, $(\delta+\delta_{10}+\delta_{15})$, reaches EIFs of only $1.52$, $1.43$, and $1.46$, which are slightly below those obtained with the single GLFD mark. This indicates that the improvement provided by the GLFD cannot probably be attributed solely to combining density information from multiple smoothing scales. 

Finally, adding the GLFD mark on top of the White mark still produces a modest but statistically significant improvement, increasing the EIF from $(2.24,\,1.87,\,1.93)$ to $(2.45,\,2.07,\,2.14)$. This suggests that the GLFD is not entirely redundant with the White mark and retains some complementary cosmological information despite their similar dependence on the local density.
Nevertheless, none of the morphology-inspired marks investigated here outperform the density-dependent White mark. This is not entirely surprising, however. Although gravitational collapse is governed by the full Hessian of the gravitational potential, the overdensity invariant (i.e. its trace) has been found to dominate the impact of the local environment on structure formation \cite{1406.4159,1801.04878,2011.10577}, with metrics of local anisotropy playing a secondary (albeit relevant) role. This is in agreement with the results found here, where the shear and morphology-inspired marks achieve modest but not insignificant improvements when combined with density-dependent mark functions. 

\section{Conclusions}\label{sec:conclusions}

Marked statistics have emerged as one of the most successful approaches for recovering cosmological information beyond that contained in the standard matter power spectrum. Almost all existing marked statistics, however, construct the mark exclusively as a function of the local density field. In this work we have explored a broader framework in which the mark may depend on any scalar quantity describing the local morphology of the cosmic web.

Our starting point was the observation that the tidal tensor provides the natural local descriptor of anisotropic gravitational collapse. Since all scalar information contained in the tidal tensor can be encoded in the triplet $(\delta,s^2,s^3)$, these quantities provide a convenient basis for constructing morphology dependent marks. We also considered the Local Fractal Dimension (LFD), which offers a complementary geometric characterisation of the cosmic web based on the scaling of the enclosed mass rather than on the local tidal field.

Within this framework we introduced several physically motivated families of marks, including the shear amplitude, isotropy mark, filament mark, Gaussian shear mark and Gaussian Local Fractal Dimension marks. Their constraining power was quantified using Fisher forecasts based on suites of Quijote $N$-body simulations for the cosmological parameters $(\Omega_m,\sigma_8,M_\nu)$.

Our first result is that the shear field itself constitutes a moderately effective marked field. The relative performance of the shear and smoothed-density marks depends on the smoothing scale: at the smallest scale explored ($R=5\,h^{-1}\mathrm{Mpc}$) the shear mark consistently outperforms the density mark for all three cosmological parameters, whereas for larger smoothing scales the density mark becomes the stronger individual observable. This behaviour suggests that the tidal anisotropy retains a larger fraction of the non-linear information present on small scales, while density-based weighting becomes increasingly effective after smoothing over larger environments. More importantly, at every smoothing scale considered, combining the density and shear marks yields substantially larger Error Improvement Factors than either observable alone. This demonstrates that the two marks probe complementary aspects of non-linear structure formation, with the anisotropy of the local gravitational field providing cosmological information that is only partially captured by density-based marked statistics.

We then investigated more elaborate morphology-inspired marks. Among those considered, none surpassed the performance of the standard White mark when used individually. This is interesting on its own, as the White mark was originally introduced largely from empirical considerations, whereas the marks proposed here were motivated by the physical morphology of the cosmic web. More interestingly, the improvement in constraining power remained almost flat across the different families of elaborated shear-based marks investigated, suggesting that the dominant source of cosmological information is the shear field itself, and not the specific function used to construct the mark. Moreover, the combination of these marks with the White mark yielded higher EIFs than the ones obtained from White alone, meaning that the information extracted by the White mark is not exhaustive and morphology-sensitive marks can probe complementary aspects of non-linear structure formation. This result is in agreement with expectations based on the relative importance of the environmental density, as opposed to other scalar functions of the tidal tensor, in determining the formation of structure \cite{1406.4159,1801.04878,2011.10577}.

We also explored the Local Fractal Dimension as an alternative metric for environmental morphology distinct from the shear field. Although conceptually different from $s$, it proved to be an effective morphology-sensitive summary of the density field. While it did not surpass the empirically-optimized White mark on its own, it produced the second largest EIFs out of all single marks considered in this work. When combined with the White mark, the joint analysis produced a modest, although tangible, improvement in constraining power. While this confirms that the GLFD captures some complementary information, the gain is not as large as one might initially expect from the individual EIFs. This restricted potential likely stems from the fact that both marks share a similar qualitative dependence on the local density: both upweight underdense environments, which inevitably introduces a degree of correlation between them. Finally, as a single mark it outperformed combinations of multiple smoothed density fields. This could suggest that the non-linear scaling of enclosed mass compresses environmental information more efficiently than simply combining density fields smoothed on different scales.

An important conceptual remark of this work is that marked statistics should be viewed not as density-weighting schemes, but more generally as local non-linear transformations of the density field designed to reorganize cosmological information before computing two-point statistics. Since both the tidal tensor and the LFD are deterministic functionals of the density field, the improvement in constraining power obtained here does not arise from introducing new information, but rather from encoding existing information in a way that becomes more accessible to well developed power spectrum estimators.

The present work represents only a first exploration of morphology-based marked statistics. We restricted our analysis to a limited set of mark families and to functions primarily dependent on the density and shear invariant $s^2$. The space of possible morphology-dependent marks is much broader and includes combinations involving the cubic shear invariant, explicit functions of the tidal eigenvalues, multiscale descriptors and perhaps data-driven optimized marks. It will also be interesting (and necessary) to investigate these methods in redshift space, with galaxy catalogues that include realistic galaxy biasing and observational systematics.

Overall, our results demonstrate that incorporating information about the morphology of the cosmic web provides a promising direction for extending marked statistics beyond purely density-based constructions.  While an optimized density-dependent will likely achieve the strongest performance individually, the framework developed in this work opens the door to a much broader class of morphology-aware statistics capable of extracting additional cosmological information from the non-linear large-scale structure of the Universe.

 \appendix
 \section{Shear Invariants in Fourier Space}\label{app:fourier}

According to \Cref{eqn:tidal_tensor_def} the tidal tensor is linear in the density field. Since the quadratic and cubic shear invariants $s^2$ and $s^3$ are local products in configuration space, their Fourier transforms are given by convolution integrals. The quadratic shear invariant can be written as
\begin{eqnarray}
    s_R^2(\bm{k})&=&\int\frac{d^3\bm{q}}{(2\pi)^3}\left(\frac{(\bm{k}-\bm{q})_i(\bm{k}-\bm{q})_j}{|\bm{k}-\bm{q}|^2}-\frac{1}{3}\delta_{ij}\right)\left(\frac{{q}^i{q}^j}{q^2}-\frac{1}{3}\delta^{ij}\right)\delta_R(\bm{k}-\bm{q})\delta_R(\bm{q})\nonumber \\
    &=&\frac{2}{3}\int\frac{d^3\bm{q}}{(2\pi)^3}P_2\left(\hat{\bm{p}}\cdot\hat{\bm{q}}\right)\delta_R(\bm{q})\delta_R(\bm{p}) 
\end{eqnarray}
where $P_2$ denotes the second Legendre polynomial. The quadratic shear invariant therefore isolates the quadrupolar component of the mode coupling, making it sensitive to anisotropic structure formation. Likewise, the cubic invariant is 
\begin{equation}
    s_R^3(\bm{k}) = \int \frac{d^3\bm{q}_1}{(2\pi)^3} \int \frac{d^3\bm{q}_2}{(2\pi)^3}\,\mathcal{K}_3(\hat{\bm{q}}_1, \hat{\bm{q}}_2, \hat{\bm{p}})\,\delta_R(\bm{q}_1)\delta_R(\bm{q}_2)\delta_R(\bm{p}) 
\end{equation}
where $\mathcal{K}_3(\hat{\bm{q}}_1, \hat{\bm{q}}_2, \hat{\bm{p}})=\left[ \mu_{12}\mu_{2p}\mu_{p1} - \frac{1}{3} \left( \mu_{12}^2 + \mu_{2p}^2 + \mu_{p1}^2 \right) + \frac{2}{9} \right]$ and $\mu_{ij}\equiv\hat{\mathbf{q}}_i\cdot\hat{\mathbf{q}}_j\,$. These expressions make explicit that both shear invariants are non-linear functionals of the density field. The quadratic invariant first appears as a second-order mode coupling with a purely quadrupolar kernel, whereas the cubic invariant involves a three-mode coupling with a kernel depending only on the relative orientations of the participating wavevectors. These kernels are closely related to the non-local operators that appear in perturbative bias expansions and effective field theory descriptions of large-scale structure.

\subsection{Perturbative Interpretation of the Shear Mark}

Here we give a short heuristic description of the perturbative structure of the power spectra involving the marked field $\tau$ associated to the mark $s_R(\bm x)$. Since $s_R$ is a non-linear function of the density field, it is intrinsically non-Gaussian even when $\delta$ is Gaussian. For $k\neq0$, the corresponding marked field satisfies
\begin{equation}
\tau(\bm{k})=s_R(\bm{k})+[s_R\,\delta](\bm{k})\,,
\end{equation}
so that
\begin{equation}
P_{\tau\tau}=P_{ss}+2P_{s,s\delta}+P_{s\delta,s\delta}\,.
\end{equation}
Some intuition can be obtained by expanding the shear amplitude around its mean value. Statistical isotropy implies
\begin{equation}
\langle s_R^2\rangle=\frac23\sigma_R^2,\hspace{12pt}{\rm where}\hspace{12pt}
\sigma_R^2=\int\frac{d^3k}{(2\pi)^3}P(k)\,W_R^2(k)
\end{equation}
is the variance of the smoothed density field. Writing
\begin{equation}
s_R^2(\bm{x})=\frac23\sigma_R^2+\Delta s_R^2(\bm{x}),
\end{equation}
a Taylor expansion gives
\begin{equation}
\begin{aligned}
s_R(\bm{x})&=
\sqrt{\frac23}\sigma_R+\sqrt{\frac38}\frac{\Delta s_R^2(\bm{x})}{\sigma_R}
-\frac{3\sqrt6}{64}\frac{\left(\Delta s_R^2(\bm{x})\right)^2}{\sigma_R^3}
+\mathcal O\!\left((\Delta s_R^2)^3\right).
\end{aligned}
\end{equation}
Substituting this expression into the definition of the marked field and removing the constant contribution yields
\begin{equation}\label{eq:tau_perturbative}
\begin{aligned}
\tau\simeq&\sqrt{\frac23}\sigma_R\,\delta+\sqrt{\frac38}\frac{\Delta s_R^2}{\sigma_R}
+\sqrt{\frac38}\frac{\delta\,\Delta s_R^2}{\sigma_R}\\
&-\frac{3\sqrt6}{64}\frac{(\Delta s_R^2)^2}{\sigma_R^3}+\cdots .
\end{aligned}
\end{equation}
\Cref{eq:tau_perturbative} shows how marked statistics reorganize the perturbative hierarchy. The cross-spectrum with the density is
\begin{equation}
\begin{aligned}
P_{\tau\delta}\simeq&\sqrt{\frac23}\sigma_R\,P_{\delta\delta}+\sqrt{\frac38}\frac{1}{\sigma_R}P_{\Delta s^2,\delta}\\
&+\sqrt{\frac38}\frac{1}{\sigma_R}P_{\delta\Delta s^2,\delta}+\cdots,
\end{aligned}
\end{equation}
where $\Delta s^2$ is itself quadratic in the density field. Therefore,
\begin{equation}
P_{\Delta s^2,\delta}\sim P_{\delta^2,\delta}
\end{equation}
is directly related to squeezed bispectrum configurations, while
\begin{equation}
P_{\delta\Delta s^2,\delta}\sim P_{\delta^3,\delta}
\end{equation}
receives leading contributions from the trispectrum.

A similar argument applies to the auto-spectrum $P_{\tau\tau}$, but is not discussed here. Developing this framework into a complete perturbative description of morphology-based marked statistics is left for future work.

\section{Detailed Results for the Shear Mark}\label{app:shear_results}

\begin{figure}[htbp]
    \centering
    \includegraphics[width=\linewidth]{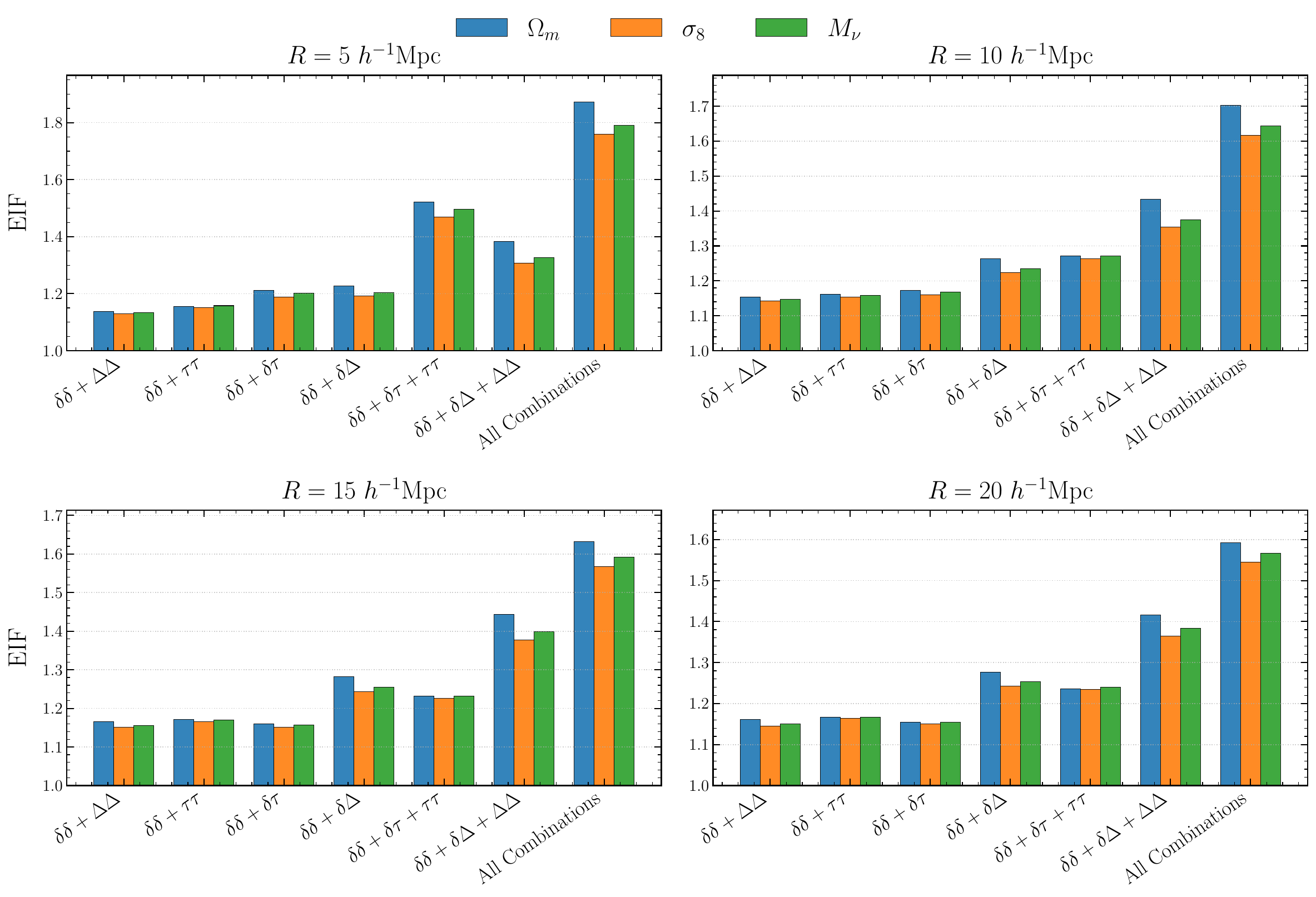}
    \caption{Same as \Cref{fig:barchart_EIF}, but showing the Error Improvement Factors (EIFs) for the four smoothing scales considered in this work, $R=5$, $10$, $15$, and $20\,h^{-1}\mathrm{Mpc}$.}
    \label{fig:big_barchart}
\end{figure}

For completeness, we report the EIFs obtained for the density, shear, and density+shear marks for all smoothing scales considered in this work. \Cref{fig:big_barchart} extends \Cref{fig:barchart_EIF} by showing the same set of marked-spectrum combinations for $R=5$, $10$, $15$, and $20\,h^{-1}\mathrm{Mpc}$. The corresponding numerical values are listed in \Cref{tab:big_table}.

\begin{table}[htpb]
    \centering
    \begin{tabular}{ccccc}
        \toprule
        \multirow{2}{*}{\textbf{Configuration}} & \multirow{2}{*}{$\boldsymbol{R \ [\text{Mpc}/h]}$} & \multicolumn{3}{c}{\textbf{EIF}} \\
        \cmidrule(lr){3-5}
        & & $\boldsymbol{\Omega_m}$ & $\boldsymbol{\sigma_8}$ & $\boldsymbol{M_\nu}$ \\
        \midrule
        
        \multirow{4}{*}{$\delta\delta + \delta\delta_R$} 
        & $5$   & $1.23$ & $1.19$ & $1.20$ \\
        & $10$  & $1.26$ & $1.22$ & $1.24$ \\
        & $15$  & $1.28$ & $1.24$ & $1.26$ \\
        & $20$  & $1.28$ & $1.24$ & $1.25$ \\
        \midrule
        
        \multirow{4}{*}{$\delta\delta + \delta_R\delta_R$} 
        & $5$   & $1.14$ & $1.13$ & $1.13$ \\
        & $10$  & $1.15$ & $1.14$ & $1.15$ \\
        & $15$  & $1.17$ & $1.15$ & $1.16$ \\
        & $20$  & $1.16$ & $1.15$ & $1.15$ \\
        \midrule
        
        \multirow{4}{*}{$\delta\delta + \delta\delta_R+\delta_R\delta_R$} 
        & $5$   & $1.38$ & $1.31$ & $1.33$ \\
        & $10$  & $1.43$ & $1.35$ & $1.38$ \\
        & $15$  & $1.44$ & $1.38$ & $1.40$ \\
        & $20$  & $1.42$ & $1.36$ & $1.38$ \\
        \midrule

        \multirow{4}{*}{$\delta\delta + \delta s_R$} 
        & $5$   & $1.21$ & $1.19$ & $1.20$ \\
        & $10$  & $1.17$ & $1.16$ & $1.17$ \\
        & $15$  & $1.16$ & $1.15$ & $1.16$ \\
        & $20$  & $1.16$ & $1.15$ & $1.15$ \\
        \midrule

        \multirow{4}{*}{$\delta\delta + s_R s_R$} 
        & $5$   & $1.16$ & $1.15$ & $1.16$ \\
        & $10$  & $1.16$ & $1.15$ & $1.16$ \\
        & $15$  & $1.17$ & $1.17$ & $1.17$ \\
        & $20$  & $1.17$ & $1.16$ & $1.17$ \\
        \midrule   
        
        \multirow{4}{*}{$\delta\delta + \delta s_R+s_Rs_R$} 
        & $5$   & $1.52$ & $1.47$ & $1.50$ \\
        & $10$  & $1.27$ & $1.26$ & $1.27$ \\
        & $15$  & $1.23$ & $1.23$ & $1.23$ \\
        & $20$  & $1.24$ & $1.24$ & $1.24$ \\
        \midrule   
        
        \multirow{4}{*}{All} 
        & $5$   & $1.87$ & $1.76$ & $1.79$ \\
        & $10$  & $1.70$ & $1.62$ & $1.64$ \\
        & $15$  & $1.63$ & $1.57$ & $1.59$ \\
        & $20$  & $1.59$ & $1.55$ & $1.57$ \\
        \bottomrule
            
    \end{tabular}
    \caption{Numerical values of the EIFs corresponding to \Cref{fig:big_barchart}.}
    \label{tab:big_table}
\end{table}

\acknowledgments

M.M.B. would like to thank Francisco Villaescusa-Navarro for his continued support with the Quijote Binder. M.M.B. acknowledges support by the Spanish Ministry of Science, Innovation and Universities under the FPU predoctoral grant FPU22/02306. J.C.D. is supported by the Spanish Research Agency (Agencia Estatal de Investigaci\'on), the Ministerio de Ciencia, Inovaci\'on y Universidades, and the European Social Funds through grant JDC2023-052152-I, as part of the Juan de la Cierva programme. JGB acknowledges support from the Spanish Research Project PID2024-159420NB-C43 [MICINN-FEDER] and the Centro de Excelencia Severo Ochoa Programme CEX2020-001007-S at IFT. JAC is funded by a Kavli/IPMU PhD Studentship. DA acknowledges support from the Beecroft Trust.

\bibliographystyle{JHEP}
\bibliography{biblio.bib}

\end{document}